\documentclass[12pt]{article}
\usepackage{amsfonts,amssymb}

\newtheorem{theorem}{Theorem}
\newtheorem{lemma}{Lemma}
\newtheorem{remark}{Remark}
\newtheorem{proposition}{Proposition}
\newtheorem{corollary}{Corollary}
\newcommand{\ju}[4]{\mbox{$
  \left(\begin{array}{cc}{#1} & {#2}\\{#3} &{#4}  \end{array}\right)$}}

\newcommand{\R}{\mathbb{R}}
\newcommand{\T}{\mathbb{T}}

\newcommand{\C}{\mathbb{C}}
\newcommand{\M}{\mathbb{M}}
\newcommand{\K}{\mathbb{K}}

\newcommand{\J}{\mathbb{J}}
\newcommand{\Z}{\mathbb{Z}} 

\newcommand{\la}{\lambda}

\newcommand{\pa}{\partial}

\newcommand{\al}{\alpha}  
\newcommand{\be}{\begin{equation}}
\newcommand{\ee}{\end{equation}}
\newcommand{\beq}{\begin{eqnarray}}
\newcommand{\eeq}{\end{eqnarray}}
\newcommand{\beqq}{\begin{eqnarray*}}
\newcommand{\eeqq}{\end{eqnarray*}}

\newcommand{\La}{\Lambda}

\title{\sf The Camassa-Holm hierarchy,
related $N$-dimensional  integrable
  systems, and 
algebro-geometric solution on a symplectic submanifold}
\author{Darryl D. Holm$^1$, Zhijun Qiao$^{1,2}$\\
$^1$T-7 and CNLS, MS B-284, Los Alamos National Laboratory\\
Los Alamos, NM 87545, USA\\
$^2$Institute of Mathematics, Fudan University\\
 Shanghai 200433, PR
China\\
{\footnotesize  E-mails: dholm@lanl.gov \ \ \ qiao@lanl.gov}
}
\date{Original vesion April 4, 2001; the 2nd version Feb. 25, 2002;\\
the 3rd  version June 9, 2002; this version July 12, 2002}
\begin{document}
\maketitle
\begin{abstract}
This paper shows that the Camassa-Holm (CH) spectral problem
yields two different integrable hierarchies of nonlinear evolution
equations (NLEEs), one is of negative order CH hierachy while the other one
is of positive order CH hierarchy. 
The two CH hierarchies possess the zero curvature representations
 through solving a key matrix equation.
We see that the well-known CH equation
is included in the negative order CH hierarchy while the Dym type
equation is included in the positive order CH hierarchy. 
Furthermore, under two  constraint conditions between the potentials
and the eigenfunctions,  the CH spectral
problem is cast in: 
\begin{enumerate} 
\item a new Neumann-like $N$-dimensional system when it  
is restricted into a symplectic submanifold of $\R^{2N}$  which is proven to be
integrable by using the Dirac-Poisson bracket and the $r$-matrix
process; and 
\item a new Bargmann-like $N$-dimensional system when it  
is considered in the whole $\R^{2N}$  which is proven to be
integrable by using the standard Poisson bracket and the $r$-matrix
process.
\end{enumerate}
In the paper, we present two $4\times 4$ instead of $N\times N$
$r$-matrix structures.
One is for the Neumann-like system (not the peaked CH system)
related to the negative order CH hierarchy, 
while the other one is for the Bargmann-like 
system (not the peaked CH system, either) 
related to the positive order hierarchy.
The whole CH hierarchy (a integro-differential hierarchy, both positive and  negative
order) 
is shown to have the parametric
solutions which obey the corresponding constraint relation.
 In particular, the  CH equation,  constrained to some
 symplectic submanifold in $\R^{2N}$, and the  Dym type equation have
 the parametric solutions. 
Moreover, we see that the kind of  parametric solution of the  CH equation
is not gauge equivalent to the peakons. Solving the parametric
representation
of solution 
on the symplectic submanifold gives a class of new  algebro-geometric solution
of the  CH equation.
 \end{abstract}
{\bf Keywords} \ \ CH hierarchy,
   Zero curvature representation, 
 $r$-matrix structure,
    Parametric solution,  CH equation, Algebro-geometric solution.

   $ $\\
{\it AMS Subject: 35Q53; 58F07; 35Q35\\
     PACS: 03.40.Gc; 03.40Kf; 47.10.+g}

\newpage

\section{Introduction}
\setcounter{equation}{0}

The shallow water equation derived by Camassa-Holm (CH) in 1993 \cite{CH} is a new  
integrable system. This equation possesses the bi-Hamiltonian
structure, Lax pair and peakon solutions, and retains higher order
terms of derivatives in a small amplitude expansion of incompressible
Euler's equations for unidirectional motion of waves at the free
surface under the influence of gravity. In 1995
Calogero \cite{Ca1995} extended the class of mechanical system of this type.
Later, Ragnisco and Bruschi \cite{RB} and Suris \cite{Su},
showed that the CH equation yields the dynamics of the peakons
in terms of an $N$-dimensional completely integrable Hamiltonian
system. Such kind of dynamical system has Lax pair and an $N\times N$
$r$-matrix structure \cite{RB}. 
 
Recently, the algebro-geometric solution on
the CH equation and the CH hierarchy arose 
much more attraction. 
This kind of solution for most classical integrable PDEs
can be obtained by using the inverse spectral transform theory,
see Dubrovin 1981 \cite{Du1}, Ablowitz and Segur 1981 \cite{AS1},
 Novikov et al 1984 \cite{N1}, Newell 1985 \cite{New1}.
This is done usually by adopting the sprectral technique
associated with the corresponding PDE. 
 Alber and Fedorov \cite{AF1, AF2} studied the stationary 
and the time-dependent  quasi-periodic
solution for  the CH equation and Dym type equation
through using the method of trace formula \cite{AC1}
and Abel mapping and functional analysis on the Riemann surfaces.
Later, Alber, Camassa, Fedorov, Holm and Marsden \cite{AC} 
considered  the trace formula under the nonstandard Abel-Jacobi
equations
and by introducing new parameters presented
the so-called weak finite-gap piecewise-smooth solutions of the integrable
CH equation and Dym type equations. Very recently, Gesztesy and Holden
\cite{GH1} discussed the algebro-geometric solutions for the CH
hierarchy using the polynomial recursion formalism and the trace
formula,
 and connected
a Riccati equation to the Lax pair of the CH equation. 

The present paper is providing another approach
to algebro-geometric solutions of the CH equation
which is constrained to some symplectic submanifold. Our approach
differs from the ones pursued in Refs. \cite{AC,AC1,AF1,AF2,GH1} and
we will outline the differences next.
Based on the nonlinearization technique \cite{Cao1989}, we 
constrain the CH hierarchy to some symplectic submanifold and use the
constraint between the potentials and the eigenfunctions 
first to give the parametric solution and then 
to give the algebro-geometric solution of the CH equation on the symplectic submanifold.

The main results of this paper are twofolds. 
\begin{itemize}
\item First, we extend
the CH equation to the negative order CH hierarchy,
which is a hierarchy of integrable
integro-differetial equations, through constructing the inverse recursion operator. 
This hierarchy is proven to have Lax pair 
through solving a key matrix equation.
The CH spectral problem 
associated with this hierarchy 
is constrained to a symplectic submanifold
and naturally gives a constraint 
between the spectral function and the potential. 
Under this constraint, the CH spectral problem (linear problem)
is nonlinearized as a new $N$-dimensional canonical Hamiltonian system
of Neumann type.  This $N$-dimensional Neumann-like system is not
 the peaked
dynamical system of the CH equation because the peakons does not come 
from the CH spectral problem.
The Neumann-like CH system is shown integrable
by using the so-called Dirac-Poisson brackets on
the symplectic submanifold in  $\R^{2N}$ and $r$-matrix process. 
Here we present a  $4\times4$ $r$-matrix
structure for the Neumann-like system, 
which is available to get the algebro-geometric solution
of the CH equation on this symplectic submanifold. 
The negative order CH hierarchy is proven
to have the parametric solution through employing the Neumann-like
constraint relation. This parametric solution does not contain
the peakons \cite{CH}, and vice versa. Furthermore, solving the
parametric representation of solution  
on the symplectic submanifold gives an algebro-geometric solution
for the CH equation. We point out that our algebro-geometric solution 
(see Eq.  (\ref{CHsolution}) and Remarks 3 and 4)
 is  different from the ones in Refs. \cite{AC,AC1,AF1,AF2,GH1},
and more simpler in the form.
 
\item Second, based on the negative case, we naturally
give the positive order CH hierarchy by considering the
recursion operator. This hierarchy is shown integrable also by solving
the same key matrix equation. The CH spectral problem, related to this
 positive order CH hierarchy,  yields a new
integrable $N$-dimensional system of Bargmann type (instead of Neumann
type) 
by using the standard Poisson bracket and $r$-matrix procedure 
in $\R^{2N}$. A $4\times4$ $r$-matrix
structure is also presented for the Bargmann-like system (not peaked CH system, either), 
which is available to get the parametric solution
of a Dym type equation  contained in the positive
order CH hierarchy. This hierarchy also possesses
the parametric solution through using the Bargmann constraint
relation.
\end{itemize}
Roughly speaking,     
our method works in the following  steps (also see \cite{Q1}):
%\begin{itemize}

 --  Start from the spectral problem.

--  Find some constraint condition between the potentials and the eigenfunctions.
       Here, for the negative CH hierarchy, we restrict it to a 
       symplectic manifold in $\R^{2N}$, but for the positive  CH hierarchy, 
       we will have the constraint condition in  the whole $\R^{2N}$.  

 --   Prove the constrained SPECTRAL PROBLEM is finite-dimenaional
       integrable. Usually we use Lax matrix and $r$-matrix procedure.

--  Verify the above constrained potential(s) is (are) a parametric solution of
       the hierarchy.

--   Solve the  parametric representation of solution in an explicit form, then give
     the  algebro-geometric solutions of the equations on the symplectic manifold.
     In this process, we  separate the variables
     of the Jacobi-Hamiltonian system \cite{SK},  then construct
the actional variables and angle-coordinates on the symplectic submanifold,
and the residues at two infinity points for some composed Riemann-Theta functions
give the algebro-geometric solutions. 
%\end{itemize}

The whole paper is organized as follows. 
Next section gives a general structure of the zero curvature
representations of the all vector fields for a given isospectral  problem.
The key point is to construct
a key matrix equation. In section 3, we present the 
negative order CH hierarchy based on the 
inverse recursion operator. 
The well-known CH equation
is included in the negative order hierarchy, and the CH spectral
problem yields a new Neumann-like system 
which is constrained to a symplectic manifold. This system 
has the canonical form and is
integrable by using 
the Dirac-Poisson bracket and the $r$-matrix process.
Here we obtain a $4\times 4$  $r$-matrix structure for the
Neumann-like CH system. 
Furthermore, 
 the whole negative order CH hierarchy constrained  
on the symplectic  submanifold has the parametric
solution. In particular, the CH equation has the parametric 
solution on this submanifold.
Finally we give an  algebro-geometric solution
of the  CH equation on this submanifold. 
In section 4, we deal with the positive order integrable 
CH hierarchy
and give a new Bargmann-like integrable system. By the use of the similar
process to section 3,
 the CH spectral
problem is nonlinearized to be an integrable system
under a Bargmann constraint.
This integrable Bargmann  system also has an $r$-matrix structure of $4\times 4$.
Moreover, the positive order CH hierarchy is also shown to have the parametric
solution which obeys the Bargmann constraint relation. 
In particular, the Dym type equation in the positive order CH hierarchy 
 has the parametric solution.

Let us now give some  symbols and convention in this paper as follows:
\begin{eqnarray*}
f^{(k)}=\left\{\begin{array}{ll}
               \frac{\pa^k}{\pa x^k}f=f_{kx}, & k=0,1,2,...,\\
               \underbrace{\int\ldots\int}_{-k}f {\rm d} x, & k=-1,-2,...,
               \end{array}
         \right.
\end{eqnarray*}
$f_t=\frac{\pa f}{\pa t}$, $f_{kxt}=\frac{\pa^{k+1} f}{\pa t\pa x^k}\
(k=0,1,2,...)$,
$\pa=\frac{\pa}{\pa x}$, $\pa^{-1}$ is the inverse of $\pa$, i.e.
$\pa \pa^{-1}=\pa^{-1}\pa=1$,
$\pa^kf$ means the operator $\pa^kf$ acts on some function $g$,
i.e.
\begin{eqnarray*}
\pa^kf\cdot g=\pa^k(fg)
=\left\{\begin{array}{ll}
               \frac{\pa^k}{\pa x^k}(fg)=(fg)_{kx}, & k=0,1,2...,\\
               \underbrace{\int\ldots\int}_{-k}fg {\rm d} x, & k=-1,-2,....
               \end{array}
         \right.
\end{eqnarray*}

In the following
the function $m$
stands for potential,
$\la$ is assumed to be a spectral parameter,
and the domain of the spatial variable
$x$ is $\Omega$ which becomes equal to
$(-\infty,\ +\infty)$ or $(0, \ T)$, while
the domain of the time variable $t_k$
is the positive time axis $\R^{+}=\left\{t_k\right|\ t_k\in \R,
\ t_k\geq 0, \ k=0,\pm1,\pm2,...\}$.
In the case $\Omega=(-\infty,\ +\infty )$,
the decaying condition at infinity
and in the case $\Omega=(0,\ T)$,
the periodicity condition for the potential function
is imposed.

    $(\R^{2N},{\rm d} p\wedge {\rm d} q)$ stands for the standard symplectic structure
in Euclid space $\R^{2N}=\left\{\left.(p,q)\right|
\ p=(p_1,\ldots,p_N), q=(q_1,\ldots,q_N)\right\}$,
$p_j,q_j$ $(j=1,\ldots,N)$ are $N$ pairs of canonical coordinates,
$\left<\cdot,\cdot\right>$ is the standard inner product in $\R^N$;
in $(\R^{2N},{\rm d} p\wedge {\rm d} q)$, the Poisson
bracket of two Hamiltonian functions
$F,H$ is defined by \cite{AV}
\begin{equation}
 \left\{F,H\right\}=\sum_{j=1}^N\left(\frac{\partial F}{\partial q_j}\frac{\partial H}{\partial p_j}
 -\frac{\partial F}{\partial p_j}\frac{\partial H}{\partial  q_j}\right)=\left<\frac{\partial F}{\partial q},\frac{\partial H}{\partial p}\right>
 -\left<\frac{\partial F}{\partial p},\frac{\partial H}{\partial  q}\right>. \label{PB}
\end{equation}
$\la_1,...,\la_N$ are assumed to be $N$ distinct spectral parameters,
 $\La=diag(\lambda _1,...,\lambda _N)$, and $I_{2\times2}=diag(1,1)$.
Denote all infinitely times differentiable functions on real field $\R$
and all integers by $C^{\infty}(\R)$ and by $\Z$, respectively.

\section{The Camassa-Holm (CH) spectral problem  
         and zero curvature  representation}
\setcounter{equation}{0}
Let us consider the Camassa-Holm (CH) spectral problem \cite{CH}:
\beq
\psi_{xx}=\frac{1}{4}\psi-\frac{1}{2}m\la\psi  \label{CH1}
\eeq
with the potential function $m$. 

Eq. (\ref{CH1}) is apparently equivalent to
\beq
y_x=Uy, \ \ U=U(m,\la)=\ju{0}{1}{\frac{1}{4}-\frac{1}{2}m\la}{0},
 \label{CHL1}
\eeq
where $y=(y_1,y_2)^T=(\psi,\psi_x)^T$.
Easy to see Eq. (\ref{CHL1})'s spectral gradient 
\be\nabla\la\equiv\frac{\delta\la}{\delta m}
=\la y_1^2\ee
satisfies the following Lenard eigenvalue problem
\be
  K\cdot\nabla\la=\la J\cdot\nabla\la
  \label{6.4.2}
\ee
with the pair of Lenard's operators 
\beq
        K=-\pa^3+\pa, \ \ \
        J =\pa m+m\pa.
         \label{6.4.3}
\eeq
They yield the recursion operator
\be
{\cal L}=J^{-1}K= (\pa m+m\pa)^{-1}(\pa-\pa^3).
              \label{6.4.4} \ee
which also has the product form ${\cal L}=\frac{1}{2}m^{-\frac{1}{2}}\pa^{-1} m^{-\frac{1}{2}}(\pa- \pa^{3})$.

Apparently, the Gateaux derivative matrix  $U_{*}(\xi)$ of the
spectral matrix  $U$  in the direction
$\xi\in C^{\infty}(\R)$ at point $m$ is
\be
U_{*}(\xi)\stackrel{\triangle}{=}\left.\frac{{\rm d}}{{\rm d} \epsilon}
\right|_{\epsilon=0}U(m+\epsilon\xi)
=\ju{0}{0}{-\frac{1}{2}\la\xi}{0} \label{6.4.5}
\ee
which is obviously an injective homomorphism.
    
   For any given $C^{\infty}$-function $G$, we construct
    the following matrix  equation with respect to    $V=V(G)$
    \beq   V_x- [U,V]=U_{*}(K\cdot G-\la J\cdot G). \label{VLCH}
    \eeq

  \begin{theorem}
For the CH spectral problem (\ref{CHL1}) and an arbitrary
$C^{\infty}$-function $G$, the matrix equation
(\ref{VLCH})  has the
following solution
\beq
  V=V(G)=\la\ju{-\frac{1}{2}G_x}{-G}{\frac{1}{2}G_{xx}-\frac{1}{4}G+\frac{1}{
2}m\la G}{\frac{1}{2}G_x}. \label{6.4.15}
\eeq
\label{Th1}
\end{theorem}

{\bf Proof}: \ \ Directly substituting Eqs. (\ref{6.4.15}),
(\ref{6.4.3}) and (\ref{6.4.5}) into Eq. (\ref{VLCH}), we
can complete the proof of this theorem.

\begin{theorem}
Let $G_0\in Ker\ J=\{G\in C^{\infty}(\R)\ |\ JG=0\}$ and
$G_{-1}\in Ker\ K=\{G\in C^{\infty}(\R)\ |\ KG=0\}$. We define
the Lenard's sequences as follows
\beq
G_j= {\cal L}^j\cdot G_0 ={\cal L}^{j+1}\cdot G_{-1}, \ j\in \Z. \label{Gj}
\eeq
Then, 
\begin{enumerate}
\item 
  the all vector fields $X_k=J\cdot G_k, \ k\in \Z$ satisfy the
following commutator representation
\beq
V_{k,x}-[U,V_k]=U_*(X_k), \ \forall k\in \Z; \label{UV1}
\eeq
\item  the following hierarchy of nonlinear evolution equations
\beq
m_{t_k}= X_k=J\cdot G_k, \ \forall k \in \Z,
\eeq
possesses the zero curvature representation
\beq
U_{t_k}-V_{k,x}+[U,V_k]=0,\ \forall k\in \Z, \label{XUV}
\eeq
\end{enumerate}
where 
\beq
V_k&=&\sum V_j\la^{k-j-1}, \ \ \sum\ = \
\left\{\begin{array}{ll}
\sum^{k-1}_{j=0}, & k>0,\\
0, & k=0,\\
-\sum^{-1}_{j=k}, & k<0,
\end{array}\right.  
\eeq
and $V_j=V(G_j)$ is given by Eq. (\ref{6.4.15}) with $G=G_j$.
\label{Th12}
\end{theorem}

{\bf Proof:} \begin{enumerate}
\item  For $k=0$, it is obvious.
For $k<0$, we have
\beqq
V_{k,x}-[U,V_k]&=&-\sum_{j=k}^{-1}\left(V_{j,x}-[U,V_j]\right)\la^{k-j-1}\\
&=& -\sum_{j=k}^{-1}U_*\left(K\cdot G_j-\la K\cdot G_{j-1}\right)\la^{k-j-1}\\
&=& U_*\left(\sum_{j=k}^{-1} K\cdot G_{j-1}\la^{k-j}-K\cdot
G_{j}\la^{k-j-1}\right) \\
&=& U_*\left( K\cdot G_{k-1}-K\cdot
G_{-1}\la^{k}\right) \\
&=& U_*( K\cdot G_{k-1})\\
&=& U_*(X_k).
\eeqq 

For the case of $k>0$, it is similar to prove.

\item  Noticing $U_{t_k}=U_*(m_{t_k})$, we obtain
\beqq
U_{t_k}-V_{k,x}+[U,V_k]= U_*(m_{t_k}-X_k).
\eeqq
The injectiveness of $U_*$ implies item 2 holds.

\end{enumerate}

\section{Negative order CH hierarchy,  integrable Neumann-like system
and algebro-geometric solution}
\setcounter{equation}{0}

\subsection{Negative order CH hierarchy}
Let us first give the negative order hierarchy
of the CH spectral problem (\ref{CHL1})
through considering the kernel element of Lenard's operator $K$.
 The kernel of operator $K$
 has the following three seed functions:
\beq
 G^1_{-1} & = &1, \label{6.4.34}\\
   G^2_{-1} & = & e^x,    \label{6.4.35}\\
 G^3_{-1} & = &e^{-x}, \label{6.4.353}
\eeq
whose all possible linear combinations form
the whole kernel of $K$.
Let $G_{-1}\in Ker \ K$, then
\be
G_{-1}=\sum_{l=1}^3a_lG^l_{-1} \label{G-1}
\ee
where $a_l=a_l(t_n), \ l=1,2,3,$
are three arbitrarily given $C^{\infty}$-functions
with respect to the time variables $t_n \ (n<0,n\in \Z)$, but independent of the spacial variable $x$.
Therefore, $G_{-1}$ directly generates an 
isospectral ($\la_{t_k}=0,\ k<0, k\in\Z$)  hierarchy
of nonlinear evolution equations for the CH spectral problem (\ref{CHL1})
   \beq 
m_{t_k} =  J{\cal L}^{k+1}\cdot G_{-1},
   \ \ \ k<0, \ k\in \Z,
   \label{6.4.36}
   \eeq
{\bf which is called the negative order CH hierarchy} because of $k<0$.
In Eq. (\ref{6.4.36}), the operator $J$ 
is defined by Eq. (\ref{6.4.3})
and ${\cal L}^{-1}$ is given by 
\beq
{\cal L}^{-1} = K^{-1}J=\pa^{-1}e^{x}\pa^{-1}e^{-2x}\pa^{-1}e^{x}
(\pa m+m\pa).      \label{6.4.9}
\eeq
Here, 
\beq
 K^{-1}  =  \pa^{-1}e^{x}\pa^{-1}e^{-2x}\pa^{-1}e^{x}.
             \label{6.4.8}
\eeq
With setting $m=u-u_{xx}$, we obtain another form of ${\cal L}^{-1}$:
\beq
{\cal L}^{-1} & = & u+ e^{x}\pa^{-1}e^{-2x}\pa^{-1}e^{x}\left(
u\pa+2u_x+\pa^{-1}m\right)\pa.       \label{6.4.91}
\eeq

By Theorem \ref{Th12}, the negative CH hierarchy (\ref{6.4.36})
 has the zero curvature representation
\beq
U_{t_k}-V_{k,x}+[U,V_k]&=&0,\ k<0, \  k\in \Z, \label{NUV}\\
V_k&=&-\sum_{j=k}^{-1} V_j\la^{k-j-1},
\eeq
i.e.
\beq 
\left\{\begin{array}{l}
y_{x}=\ju{0}{1}{\frac{1}{4}-\frac{1}{2}m\la}{0}y,\\
y_{t_k}=-\sum_{j=k}^{-1}
\ju{-\frac{1}{2}G_{j,x}}{-G_j}{\frac{1}{2}G_{j,xx}-\frac{1}{4}G_j+\frac{1}{
2}m\la G_j}{\frac{1}{2}G_{j,x}}\la^{k-j}y, \\
\ \ \ \ \ \ \ \ \ \  k=-1,-2,...,
\end{array}\right. \label{Laxch}
\eeq
where 
$G_j={\cal L}^{j+1}\cdot G_{-1}, \ j<0,\ j\in \Z$.
Thus, all nonlinear equations in the negative CH hierarchy (\ref{6.4.36}) are
integrable.

Let us now give some special reductions of Eq. (\ref{6.4.36}).
\begin{itemize}
\item
In the case of $a_1=-1, a_2=a_3=0$, i.e. $G_{-1}=-G^1_{-1}=-1$,
because $G_{-2}={\cal L}^{-1}\cdot G_{-1}=-u$, the second
equation of Eq. (\ref{6.4.36}) reads
\end{itemize}
\beq
 m_{t_{-2}}  = -(\pa m+m\pa)\cdot u, \label{6.4.37}
\eeq
i.e. (here noticing $m=u-u_{xx}$)
\beq 
u_{t_{-2}}-u_{xx,t_{-2}}+3uu_x=2u_xu_{xx}+uu_{xxx}
 \label{6.4.39}
\eeq
which is exactly the Camassa-Holm equation \cite{CH}. According to
Eq. (\ref{Laxch}), the CH equation (\ref{6.4.39}) possesses the
following zero curvature representation
\beq 
\left\{\begin{array}{l}
y_{x}=\ju{0}{1}{\frac{1}{4}-\frac{1}{2}m\la}{0}y,\\
y_{t_{-2}}=
\ju{-\frac{1}{2}u_{x}}{-u-\la^{-1}}{\frac{1}{2}mu\la+\frac{1}{4}u-
\frac{1}{
4}\la^{-1}}{\frac{1}{2}u_{x}}y,
\end{array}\right. \label{Laxch2}
\eeq
which is equivalent to 
\beq
\left\{\begin{array}{l}
\psi_{xx}=\frac{1}{4}\psi-\frac{1}{2}m\la\psi,\\
\psi_{t_{-2}}=\frac{1}{2}u_{x}\psi-u\psi_x-\la^{-1}\psi_x.
\end{array}\right. \label{Laxch11}
\eeq
Eq. (\ref{Laxch11}) is coinciding with the one in Ref. \cite{CH}.

\begin{itemize}
\item
In the cases of $a_1=0, a_2=1,a_3=0$ and
$a_1=0,a_2=0,a_3=1$, i.e. $G_{-1}=e^x, e^{-x}$,
we can write them in a uniform expression:
$$G_{-1}=e^{\epsilon x}, \ \ \epsilon=\pm1.$$
\end{itemize}
The first equation of Eq. (\ref{6.4.36}) reads
\beq
m_{t_{-1}}=(m_x+2\epsilon m)e^{\epsilon x}
\eeq
which is a linear PDE.

Because $G_{-2}={\cal L}^{-1}\cdot G_{-1}=
\left(u+\epsilon u^{(-1)}\right)e^{\epsilon x}$, 
the second equation of Eq. (\ref{6.4.36}) reads
\beq 
m_{t_{-2}}=\left(m_x\left(u+\epsilon u^{(-1)}\right)+2m
\left(u_x+2\epsilon u+u^{(-1)}\right)\right)e^{\epsilon x}
 \label{6.4.39m}
\eeq
where $m=u-u_{xx}$.
This equation has the following zero curvature representation
\beq 
\left\{\begin{array}{l}
y_{x}=\ju{0}{1}{\frac{1}{4}-\frac{1}{2}m\la}{0}y,\\
y_{t_{-2}}=V_{-2}y,\\
\end{array}\right. \label{Laxch3}
\eeq
where {\footnotesize
\beqq
V_{-2}&=&-V(G_{-2})\la^{-1}-V(G_{-1})\la^{-2}\\
& =&e^{\epsilon x}\ju
{\frac{1}{2}(u_{x}+2\epsilon u+u^{(-1)})+\epsilon \la^{-1}}
{u+\epsilon u^{(-1)})+ \la^{-1}}
{\frac{3}{2}\epsilon u_x+\frac{7}{4}u+\frac{3}{4}\epsilon u^{(-1)}
+\frac{1}{2}m\left(u+\epsilon u^{(-1)}\right)\la+\frac{1}{4}\la^{-1}
}
{-\frac{1}{2}\left(u_{x}+2\epsilon u+u^{(-1)}\right)+\epsilon \la^{-1}}.
\eeqq}
Eq. (\ref{Laxch3})
can be changed to the following Lax form 
\beq
\left\{\begin{array}{l}
\psi_{xx}=\frac{1}{4}\psi-\frac{1}{2}m\la\psi\\
\psi_{t_{-2}}=\left(u+\epsilon u^{(-1)}+\la^{-1}\right)e^{\epsilon x}
\psi_x-\frac{1}{2}\left(u_x+2\epsilon u+u^{(-1)}+\la^{-1}\right)e^{\epsilon x}
\psi.
\end{array}\right. \label{Laxch4}
\eeq
Both of the two cases: $\epsilon =\pm1$ for Eq. (\ref{6.4.39m}) are integrable.

\subsection{$r$-matrix structure for the Neumann-like CH system}

Consider the following matrix  (called `negative' Lax matrix)
\begin{equation}
L_-\left(\lambda\right)=\left(\begin{array}{lr}
A_-\left(\lambda\right) & B_- \left(\lambda\right)\\
C_-\left(\lambda\right) & -A_-\left(\lambda\right)
\end{array}\right)  \label{GL-}
\end{equation}
where
\begin{eqnarray}
A_-\left(\lambda\right)&=& -\left<p,q\right>\la^{-1}+
           \sum_{j=1}^{N}\frac{p_{j}q_{j}}{\la-\la_{j}}, \\
B_-\left(\lambda\right)&=&\la^{-2}+\left<q,q\right>\la^{-1}-\sum_{j=1}^{N}\frac{q_{j}^{2}}{\la-\la_{j}},\\
C_-\left(\lambda\right)&=&\frac{1}{4}\la^{-2}-\left<p,p\right>\la^{-1}+\sum_{j=1}^{N}\frac{p_{j}^{2}}{\la-\la_{j}}.
\end{eqnarray}

We calculate the determinant of $L_-(\la)$:

\begin{eqnarray}
\frac{1}{2}\la^2\det L_-\left(\la\right)&=&-\frac{1}{4}\la^2{\rm Tr}L_-^{2}
\left(\la\right)
 =-\frac{1}{2}\la^2\left(A_-^2\left(\lambda\right)+B_-\left(\lambda\right)C_-\left(\lambda\right)\right) \nonumber\\
                &=&\sum_{j=-2}^{1}H_j\la^{j}+
                \sum_{j=1}^{N}\frac{E_{j}^-}{\la-\la_{j}}
\label{detL-}
\end{eqnarray}
where ${\rm Tr}$ stands for the trace of a matrix, and {\normalsize
\begin{eqnarray}
H_{-2}&=& -\frac{1}{8},\nonumber\\
H_{-1}&=& \frac{1}{2}\left<p,p\right>-\frac{1}{8}\left<q,q\right>,\\
H_{0}&=& \left<p,q\right>\left<\La p,q\right>-\left<p,q\right>^2,\nonumber\\
H_{1}&=& -\left<p,q\right>\left<\La^{-1} p,q\right>+ \left<p,q\right>^2,\nonumber\\
E_{j}^-&=& \left<p,q\right>\la_jp_jq_j
     -\frac{1}{2}\left(\left<q,q\right>\la_{j}+1\right)p_j^2 \nonumber\\
   & &   -\frac{1}{2}\left(\left<p,p\right>\la_{j}-\frac{1}{4}\right)q_j^2
   +\frac{1}{2}\la_j^2\Gamma_j^-, \ \ \ j=1,2,...,N,\\
   & & \Gamma_j^- =\sum_{l\not=j,l=1}^N\frac{(p_jq_l-p_lq_j)^2}{\la_j-\la_l}.
   \nonumber
\end{eqnarray}}

Let
\beq
F_k=\sum_{j=1}^{N}\la^{k+1}_jE_j^-,\ \ k=-1,-2,\ldots,
\eeq
then it reads
\beq
F_{k}&=& \frac{1}{2}\left<\La^{k+1}p, p\right>-
         \frac{1}{8}\left<\La^{k+1}q,q\right>\nonumber\\
     & &+\frac{1}{2}\sum_{j=k}^{-2}\left(
     \left<\La^{j+2}q,q\right>\left<\La^{k-j}p,p\right>-
     \left<\La^{j+2}p,q\right>\left<\La^{k-j}p,q\right>\right),\label{Fk}\\
 & & \ \ \ \ \ \ \ \ \ \ \ \  k=-1,-2,-3,.... \nonumber
\eeq
Obviously, $F_{-1}=H_{-1}$.

Now, we consider the following symplectic submanifold in $\R^{2N}$
\beq \mathbb{M}=\left\{(q,p)\in \R^{2N}\left|\ F\equiv \frac{1}{2}\left(\left<\La q,q\right>-1\right)=0,\right.
G\equiv \left<\La q,p\right>=0\right\} \label{4.2}\eeq
and  introduce the Dirac bracket  on $\mathbb{M}$
 \be \{f,g\}_D=\{f,g\}+ \frac{1}{\left<\La^2 q,q\right>}\left(\{f,F\}\{G,g\}-\{f,G\}\{F,g\}\right)
 \label{Dirac}\ee
which is easily proven to be bilinear, skew-symmetric
and satisfy the Jacobi identity.

In particular, the Hamiltonian system $(H_{-1})_D$: \ $q_x=\{q,H_{-1}\}_D,\ \
p_x=\{p,H_{-1}\}_D$ on $\mathbb{M}$ reads
\beq
(H_{-1})_D: \ \ \
\left\{\begin{array}{l}
q_x=p,\\
p_x=\frac{1}{4}q-\frac{1+4\left<\La p,p\right>}{4\left<\La^2 q,q \right>}\La q,\\
\left<\La q,p\right>=0, \ \ \left<\La q,q \right>=1.
\end{array}\right. \label{H-1D}
\eeq
We call this a Neumann-like system on $\M$. 
Let
\beq
m&=& \frac{1+4\left<\La p,p\right>}{2\left<\La^2 q,q \right>},\label{mch}\\
y_1&=&q_j, \ y_2\ = \ p, \ \ \ \la=\la_j, \ \ j=1,...,N.
\eeq
Then, the Neumann-like flow $(H_{-1})_D$ 
on $\mathbb{M}$ exactly becomes
\beq
y_x=U(m,\la)y, \ \ y=(y_1,y_2)^T,   \label{CH12}
\eeq
which is nothing else but the CH spectral problem (\ref{CHL1}) 
with the potential function $m$. Therefore, we can call
the canonical Hamiltomnian system (\ref{H-1D}) {\bf the Neumann-like
CH system on $\M$}.

A long but direct computation leads to
the following key equalities:
\beqq
\{A_-(\la),A_-(\mu)\}_D&=&\{B_-(\la),B_-(\mu)\}_D=0,\\
\{C_-(\la),C_-(\mu)\}_D&=& \frac{1+4 \left<\La p,p\right>}{\left<\La^2 q,q\right>}
\left( \frac{\la}{\mu}A_-(\la)- \frac{\mu}{\la}A_-(\mu)\right)\nonumber\\
& &+\frac{4\la\mu}{\left<\La^2 q,q\right>}
\left( C_-(\la)A_-(\mu)- C_-(\mu)A_-(\la)\right),\\
\{A_-(\la),B_-(\mu)\}_D&=&
 \frac{2}{\mu-\la}(B_-(\mu)-B_-(\la))+ \frac{2}{\la}B_-(\mu)+
      \frac{2}{\mu}B_-(\la)\nonumber\\
    & &  -\frac{2\la\mu}{\left<\La^2 q,q\right>}B_-(\la)B_-(\mu),\\
\{A_-(\la),C_-(\mu)\}_D&=&
 \frac{2}{\mu-\la}(C_-(\la)-C_-(\mu))- \frac{2}{\la}C_-(\mu)-
      \frac{2}{\mu}C_-(\la)\nonumber\\
    & &  +\frac{2\la\mu}{\left<\La^2 q,q\right>}B_-(\la)C_-(\mu)-
\frac{(1+4 \left<\La p,p\right>)\la}{2\left<\La^2 q,q\right>\mu}B_-(\la),\\
\{B_-(\la),C_-(\mu)\}_D&=&
 \frac{4}{\mu-\la}(A_-(\mu)-A_-(\la))+ \frac{4}{\la}A_-(\mu)+
      \frac{4}{\mu}A_-(\la)\nonumber\\
     & & -\frac{4\la\mu}{\left<\La^2 q,q\right>}B_-(\la)A_-(\mu).
\eeqq

$ $

Let $L^-_{1}\left(\lambda\right)=L_-\left(\lambda\right)\otimes I_{2\times2}$,
 $L^-_{2}\left(\mu\right)=I_{2\times2}\otimes L_-\left(\mu\right)$,
where $L_-\left(\la\right),\ L_-\left(\mu\right)$ are given through
Eq. (\ref{GL-}).
In the following, we search for a general $4\times 4$   $r$-matrix
structure
$r^-_{12}\left(\lambda,\mu\right)$ such that the fundamental Dirac-Poisson bracket:
\begin{equation}
\left\{L_-\left(\lambda\right) \stackrel{\otimes}{,}
L_-\left(\mu\right)\right\}_D=\left[r_{12}^-\left(\lambda,\mu\right),
L^-_{1}\left(\lambda\right)\right]-\left[r^-_{21}\left(\mu,\lambda\right),
L^-_{2}\left(\mu\right)\right]  \label{fpb-}
\end{equation}
holds, where the entries of the $4\times4$ matrix
$\left\{L_-\left(\lambda\right) \stackrel{\otimes}{,}
L_-\left(\mu\right)\right\}_D$ are
$$\left\{L_-\left(\lambda\right) \stackrel{\otimes}{,}
L_-\left(\mu\right)\right\}_{D_{kl,mn}}=\left\{L_-\left(\lambda\right)_{km},
L_-\left(\mu\right)_{ln}\right\}_D, \ \ k,l,m,n=1,2,$$
and
$r^-_{21}\left(\lambda,\mu\right)=Pr^-_{12}\left(\lambda,\mu\right)P,$ with
\beq
P&=&\frac{1}{2}\left(I_{2\times2}+\sum_{j=1}^3\sigma_j\otimes\sigma_j\right)=
 \left(\begin{array}{cccc}
1& 0 & 0 & 0\\
0& 0 & 1 & 0\\
0& 1 & 0 & 0\\
0& 0 & 0 & 1
\end{array}
\right), \nonumber
\eeq
where $\sigma_j'$ are the Pauli matrices.

$ $

   \begin{theorem}
\be
r^-_{12}\left(\lambda,\mu\right)=\frac{2\la}{\mu(\mu-\la)}P+S^- \label{r}
\ee
is an   $r$-matrix   structure satisfying Eq.  (\ref{fpb-}), where
\begin{eqnarray*}
S^-&=& \frac{\la\left(1+4 \left<\La p,p\right>\right)}{2\mu\left<\La^2 q,q\right>}
\ju{0}{0}{1}{0} \otimes \ju{0}{0}{1}{0}\nonumber\\
& & +
\frac{2\la\mu}{\left<\La^2 q,q\right>}
\ju{0}{0}{0}{1} \otimes \ju{-B_-(\la)}{0}{2A_-(\mu)}{B_-(\la)}\nonumber\\
&=& \left(\begin{array}{cccc}
0& 0 & 0 & 0\\
0& 0 & 0 & 0\\
0& 0 & -\frac{2\la\mu}{\left<\La^2 q,q\right>}B_-(\la) & 0\\
\frac{\la\left(1+4 \left<\La p,p\right>\right)}{2\mu\left<\La^2 q,q\right>}
 & 0 & \frac{4\la\mu}{\left<\La^2 q,q\right>}A_-(\mu)
 & \frac{2\la\mu}{\left<\La^2 q,q\right>}B_-(\la)
 \end{array}\right).
\end{eqnarray*}
\end{theorem}

$ $

 Apparently, our $r$-matrix
structure
 (\ref{r}) is of $4\times 4$ and is different from the one in Ref. \cite{RB}.

\subsection{Integrability}

Because there is an   $r$-matrix   structure
satisfying Eq.  (\ref{fpb-}),
\begin{equation}
\left\{L_-^2\left(\lambda\right) \stackrel{\otimes}{,} L_-^{2}
\left(\mu\right)\right\}_D=\left[\bar{r}^-_{12}\left(\lambda,\mu\right),
L^-_{1}\left(\lambda\right)\right]-
\left[\bar{r}^-_{21}\left(\mu,\lambda\right), L^-_{2}\left(\mu\right)\right]
\end{equation}
where
\beqq
\bar{r}^-_{ij}\left(\lambda,\mu\right)=\sum_{k=0}^{1}\sum_{l=0}^{1}
(L^-_{1})^{1-k}\left(\lambda\right)
(L^-_{2})^{1-l}\left(\mu\right) r^-_{ij}\left(\lambda,\mu\right)
(L^-_{1})^{k}\left(\lambda\right)
(L^-_{2})^{l}\left(\mu\right),\\
  ij=12, 21.
\eeqq
Thus,
\begin{equation}
4\left\{{\rm Tr}L_-^{2}\left(\lambda\right), {\rm Tr}L_-^{2}\left(\mu\right)
\right\}_D={\rm Tr}\left\{L_-^{2}\left(\lambda\right) \stackrel{\otimes}{,}
L_-^{2}\left(\mu\right)\right\}_D=0.
\end{equation}
So, by Eq.  (\ref{detL-}) we immediately obtain the following theorem.

\begin{theorem}
The following equalities
\begin{eqnarray}
& &\{E^-_{i}, E^-_{j}\}_D=0, \  \{H_l, E^-_{j}\}_D=0, \   \{F_k, E^-_{j}\}_D=0, \\
& & i, j=1,2,\ldots,N,\ l=-2,-1,0,1,\ k=-1,-2,\dots, \nonumber
\end{eqnarray}
hold. Hence, the Hamiltonian systems $\left(H_l\right)_D$ and $\left(F_k\right)_D$ on
$\mathbb{M}$
\begin{eqnarray}
\left(H_{l}\right)_D&:& 
\quad q_x=\{q,H_l\}_D, \   p_x=\{p, H_l\}_D, \ \  l=-2,-1,0,1, \\
\left(F_{k}\right)_D&:& \quad q_{t_{k}}=\{q, F_{k}\}_D, 
\   p_{t_{k}}=\{p,F_{k}\}_D, \quad   k=-1,-2,\ldots,
\end{eqnarray}
are completely integrable.
\end{theorem}

$ $

In particular,  we obtain the following results.
\begin{corollary}
The Hamiltonian system $(H_{-1})_D$ defined by Eq. (\ref{H-1D})
is completely integrable.
\end{corollary}

    \begin{corollary} 
 All composition functions $f\left(H_l,F_k\right)$,  $f\in
C^{\infty}\left(\R\right)$, $k=-1,-2,...,$\\
 are completely integrable Hamiltonians on $\mathbb{M}$.
\end{corollary}

\subsection{Parametric solution of the negative order CH
  hierarchy restricted onto $\M$}

In the following, we consider the relation between
constraint and nonlinear equations in the negative order
CH hierarchy (\ref{6.4.36}). Let us start from the following
setting
\beq
 G^1_{-1} & = & \sum_{j=1}^N\nabla\la_j,
  \label{4.10-}
\eeq
where $G^1_{-1}=1$, and $ \nabla\la_j=\la_jq^2_j$
is the functional gradient of the CH spectral problem
(\ref{CHL1}) corresponding to the spectral parameter $\la_j$
($j=1,...,N$).

  Apparently Eq. (\ref{4.10-}) reads
\beq
\left<\La q,q\right>=1. \label{qq1}
\eeq
After we do one time derivative with respect to $x$, we 
get 
\beq
 \left<\La p,q\right>=0,\ \ p=q_x. \label{4.13}
  \eeq
This equality together with Eq. (\ref{qq1}) forms the symplectic submanifold
$\M$ we need. Apparently, derivating Eq. (\ref{4.13}) leads to the constraint
relation (\ref{mch}).

\begin{remark}
Because $\mathbb{M}$ defined by Eq. (\ref{4.2})
is not the usual tangent bundle, i.e.
$\mathbb{M}\not=TS^{N-1}=
\left\{\left.(q,p)\in \R^{2N}\right|\
\tilde{F}\equiv \left<q,q\right>-1=0,
     \tilde{G}\equiv \left<q,p\right>=0\right\}$ and Eq. (\ref{mch})
     is not the usual Neumann constraint on $TS^{N-1}$,
Eq. (\ref{H-1D}) is therefore a new kind of Neumann system.
In subsection 3.3 we have proven its integrability.
\end{remark}

Since the Hamiltonian flows $(H_{-1})_D$ and $(F_k)_D$ on $\mathbb{M}$ are completely
integrable and their Poisson brackets $\{H_{-1},F_k\}_D=0$ ($k=-1,-2,...$),
their phase flows $g^x_{H_{-1}},\ g^{t_k}_{F_k}$ commute \cite{AV}. Thus,
we can define their compatible solution as follows:
\beq
\left(\begin{array}{l}
q(x,t_k)\\
p(x,t_k)
\end{array}
\right)=g^x_{H_{-1}} g^{t_k}_{F_k}\left(\begin{array}{l}
q(x^0,t_k^0)\\
p(x^0,t_k^0)
\end{array}
\right), \ \ k=-1,-2,...,
\eeq
where $x^0, \ t_k^0$ are the initial values of phase flows
$g^x_{H_{-1}},\ g^{t_k}_{F_k}$.

\begin{theorem}
Let $q(x,t_k), \ p(x,t_k)$ be a solution of the compatible Hamiltonian systems
$(H_{-1})_D$ and $(F_k)_D$ on $\M$. Then
\beq
m=\frac{1+4\left<\La p(x,t_k),p(x,t_k)\right>}
{2\left<\La^2 q(x,t_k),q(x,t_k)\right>} \label{mr-}
\eeq
satisfies the constrained negative order CH hierarchy
(i.e. constrained to $\M$)
   \beq
m_{t_k} & = & J{\cal L}^{k+1}\cdot 1,
   \ \  k=-1,-2,...,
   \label{umkdv-}
   \eeq
where the operators $J$ and ${\cal L}^{-1}$ are given by Eqs. (\ref{6.4.3})
and (\ref{6.4.9}), respectively.
   \label{th4-}
\end{theorem}
{\bf Proof}:  On one hand, the recursion operator ${\cal L}$
acts on Eq. (\ref{4.10-}) and results in the following
\beq
J{\cal L}^{k+1}\cdot  G^1_{-1}&=&J\cdot
          \left<\La^{k+2}q,q\right>\nonumber\\
&=& m_x\left<\La^{k+2}q,q\right>+4m\left<\La^{k+2}q,p\right>\nonumber\\
&=& \frac{2(1+4 \left<\La p,p\right>)}{\left<\La^2 q,q\right>^2}
\left(\left<\La^2 q,q\right>\left<\La^{k+2}q,p\right>-\left<\La^2 q,p\right>\left<\La^{k+2}q,q\right>\right).\nonumber\\
\eeq
In this procedure, Eqs. (\ref{6.4.2}) and (\ref{H-1D}) are used.

On the other hand, we directly make the derivative
of Eq. (\ref{mr-}) with respect to $t_k$.
Then we obtain
\beq
m_{t_k}&=& \frac{4\left<\La^2 q,q\right>\left<\La^2 p,p_{t_k}\right>-(1+4 \left<\La p,p\right>)\left<\La^2 q,q_{t_k}\right>}{ \left<\La^2 q,q\right>^2}
\label{mr1}
\eeq
where $q=q(x,t_k),\ p=p(x,t_k)$. But,
\beq
q_{t_k}&=&\{q,F_k\}_D, \ \ \ p_{t_k}=\{p, F_k\}_D,\ \ k=-1,-2,...,
\eeq
where $F_k$ are defined by Eq. (\ref{Fk}), i.e.
\beq
q_{t_k}&=&\sum_{j=k}^{-1} \left(\left<\La^{j+2} q,q\right>\La^{k-j}p
-\left<\La^{j+2} q,p\right>\La^{k-j}q \right),\\
p_{t_k}&=&\frac{1+4 \left<\La p,p\right>}{4\left<\La^2 q,q\right>}
\left(\left<\La^2 q,q\right>\La^{k+1}q-\left<\La^{k+2} q,q\right>
\La q\right)\nonumber\\
& &  \ \ +\sum_{j=k}^{-1} \left(\left<\La^{j+2} q,p\right>\La^{k-j}p
-\left<\La^{j+2} p,p\right>\La^{k-j}q \right).
\eeq
Therefore after substituting them into Eq. (\ref{mr1})
and calculating it, we have
\beqq
m_{t_k}&=&\frac{2(1+4 \left<\La p,p\right>)}{\left<\La^2 q,q\right>^2}
\left(\left<\La^2 q,q\right>\left<\La^{k+2}q,p\right>-\left<\La^2 q,p\right>\left<\La^{k+2}q,q\right>\right)
\eeqq
which completes the proof.

$ $
\begin{lemma}
Let $q, \ p$ satisfy the integrable
Hamiltonian system
$(H_{-1})_D$. Then
on the symplectic submanifold $\mathbb{M}$, we have
\begin{enumerate}
\item
\beq
\left<q,q\right>-4\left<p,p\right>=0.
\label{lemma1}
\eeq

\item
\beq
u=\left<q(x,t_k),q(x,t_k)\right>, \ \ k,=-1,-2,..,
\label{unk}
\eeq
satisfies the equation $m=u-u_{xx}$, where $m$ is given by
Eq. (\ref{mr-}), and $q(x,t_k), \ p(x,t_k)$ be a solution
of the compatible integrable Hamiltonian systems
$(H_{-1})_D$ and $(F_k)_D$ on $\M$.
\end{enumerate}
\end{lemma}
{\bf Proof:} \ \
\beq
(\left<q,q\right>-4\left<p,p\right>)_x&=&2\left<q,p\right>-8\left<p,p_x\right>
\nonumber\\
                                  &=& 0\nonumber
\eeq
completes the proof of the first equality.

As for the second one, we have
\beq
u-u_{xx}&=& \left<q,q\right>
-2\left<p,p\right>-2\left<q,\frac{1}{4}q-\frac{1}{2}m\La q\right> \nonumber\\
&=&\frac{1}{2}\left<q,q\right> -
2\left<p,p\right>+m \nonumber\\
&=&m.                         \nonumber
\eeq

In particular, with $k=-2$, we obtain the following corollary.
\begin{corollary}
Let $q(x,t_{-2}), \ p(x,t_{-2})$ be a
solution of the compatible integrable Hamiltonian systems
$(H_{-1})_D$ and $(F_{-2})_D$ on $\M$. Then
\beq
m&=&m(x,t_{-2})=-\frac{1+4\left<\La p(x,t_{-2}),p(x,t_{-2})\right>}
{2\left<\La^2 q(x,t_{-2}),q(x,t_{-2})\right>}, \label{mr2-}\\
u&=&u(x,t_{-2})=\left<q(x,t_{-2}),q(x,t_{-2})\right>,
\label{u2-}
\eeq
satisfy the CH equation (\ref{6.4.37}) on $\M$.
Therefore, $u=\left<q(x,t_{-2}),q(x,t_{-2})\right>,$ is a solution of
the  CH equation (\ref{6.4.39}) on $\M$.
Here $H_{-1}$ and $F_{-2}$ are given by
\beq
H_{-1}&=& \frac{1}{2}\left<p,p\right>-\frac{1}{8}\left<q,q\right>,\nonumber\\
F_{-2}&=&
\frac{1}{2}\left<\La^{-1}p,p\right>-
\frac{1}{8}\left<\La^{-1}q,q\right>
+\frac{1}{2}\left(\left<q,q\right>\left<p,p\right>-\left<q,p\right>^2\right).\nonumber
\eeq \label{Coro3}
\end{corollary}

{\bf Proof:} \ \ Via some direct calculations, we obtain 
\beqq
m_{t_{-2}}=-\frac{2(1+4 \left<\La p,p\right>)}{\left<\La^2 q,q\right>^2}
\left(\left<\La^2 q,q\right>\left<q,p\right>-\left<\La^2 q,p\right>\left<q,q\right>\right).
\eeqq
And lemma 1 gives
\beqq
-J\cdot u&=& -J{\cal L}^{-1}\cdot G_{-1}^1\\
&=&J\cdot <q,q>\\
&=&-\frac{2(1+4 \left<\La p,p\right>)}{\left<\La^2 q,q\right>^2}
\left(\left<\La^2 q,q\right>\left<q,p\right>-\left<\La^2
    q,p\right>\left<q,q\right>\right)\\
&=&m_{t_{-2}}
\eeqq
which completes the proof.

$ $

By Theorem \ref{th4-}, the constraint relation given by Eq. (\ref{mch})
is actually a solution of the constrained negative order CH hierarchy
(\ref{6.4.36}) on $\M$.
Thus, we also {\bf call the system $(H_{-1})_D$ (i.e. Eq. (\ref{H-1D}))
  a  negative order restricted  CH flow (Neumann-like) of the spectral problem (\ref{CHL1})
on the symplectic submanifold $\M$.  All Hamiltonian systems
$(F_k)_D, \ k<0,k\in\Z$ are therefore called the negative 
order integrable restricted flows (Neumann-type) on $\M$}.

\begin{remark}
Of course, we can also consider the integrable Bargmann-like
CH system associated with the positive order CH hierarchy (\ref{6.4.19}).
Please see this in section 4.
A systematic approach to
generate new integrable negative order
hierarchies can be seen in Ref. \cite{Q}.
\end{remark}

\subsection{Comparing parametric solution with  peakons}
  Let us now compare the Neumann-like 
CH system $(H_{-1})_D$ with the peakons dynamical system.

Let $P_j, Q_j\ (j=1,2,...,N)$ be peakons dynamical variables of the CH
equation (\ref{6.4.39}), then with \cite{CH} 
\beq
u(x,t)&=& \sum^N_{j=1}P_j(t)e^{-|x-Q_j(t)|},\label{chu1}\\
m(x,t)&=& \sum^N_{j=1}P_j(t)\delta(x-Q_j(t)),\label{chm1}
\eeq 
where $t=t_{-2}$ and $\delta(x)$ is the $\delta$-function,
 the CH equation (\ref{6.4.39}) yields a canonical peaked Hamiltonian system
\beq \left. \begin{array}{l}
\dot{Q}_j(t)=\frac{\pa  H}{\pa P_j}=\sum_{k=1}^NP_k(t)e^{-|Q_k(t)-Q_j(t)|}\\
\dot{P}_j(t)=-\frac{\pa  H}{\pa Q_j}=-P_j\sum_{k=1}^NP_k(t)sgn\left(Q_k(t)-Q_j(t)\right)
     e^{-|Q_k(t)-Q_j(t)|}
\end{array}\right. \label{PQ1}
\eeq
with 
\beq
H(t)&=&\frac{1}{2}\sum_{i,j=1}^NP_i(t)P_j(t)e^{-|Q_i(t)-Q_j(t)|}. \label{Ht1}
\eeq
In Eq. (\ref{PQ1}), ``sgn'' means the signal function. The same
meaning happens in the following.

A natual question is: what is the relationship between the peaked
Hamiltonian system (\ref{PQ1}) and the Neumann-like systems
(\ref{H-1D}) and ($F_{-2}$)? Apparently, the peaked system (\ref{PQ1})
does not include the systems (\ref{H-1D}) and ($F_{-2}$)
because the system (\ref{PQ1}) is only concerned about the time part.
   
By Corollary \ref{Coro3}, we know that
\be
u=u(x,t)=\left<q(x,t),q(x,t)\right>=\sum_{j=1}^Nq_j^2(x,t),\label{chu2}\ee
is a solution
of the  CH equation (\ref{6.4.39}) on $\M$,
where we set $t_{-2}=t$, and $q_j(x,t), \
p_j=\frac{\pa q_j(x,t)}{\pa x}$ satisfies the two
integrable commuted systems $(H_{-1})_D, \ (F_{-2})_D$
on the symplectic submanifold $\M$.   

Assume $P_j(t), \ Q_j(t)$ are the solutions of the peaked system
(\ref{PQ1}), 
we make the following transformation (when $P_j(t)<0$, we use 
$\sqrt{P_j(t)}=i\sqrt{-P_j(t)}$)
\beq
q_j(x,t)=\sqrt{P_j(t)}e^{-\frac{1}{2}|x-Q_j(t)|}, \label{qj1}
\eeq
which implies
\beq
p_j(x,t)=\frac{d }{d x}q_j(x,t)=-\frac{1}{2}sgn(x-Q_j(t))q_j. \label{pj1}
\eeq

Hence, we obtain
\beq
\frac{d }{d x}p_j(x,t)= \frac{d^2 }{d x^2}q_j(x,t)=-\delta(x-Q_j(t))q_j+\frac{1}{4}q_j.
\label{Qq1}\eeq
However, on $\M$ we have the constraints
$ <\La q, q>=1, \ <\La q,p>=0$
which implies 
\beq \sum_{j=1}^N\la_jq_j^2\delta(x-Q_j)=\frac{1}{2},\ \  \forall x\in
\R. \label{xd1} \eeq
This equality is obviously not true! So, the CH peakon system 
(\ref{PQ1}) does not coincide with the nonlinearized CH
spectral problem (\ref{H-1D}).

Let us now furthermore  compute the derivative with respect to $t$.
Inserting  the peakon system  (\ref{PQ1}), we get
\beq
& & \dot{q}_j(x,t)=\frac{d }{d t}q_j(x,t)\nonumber\\
&=& \frac{1}{2}q_j\sum_{k=1}^NP_k(t)\Big[sgn(x-Q_j(t))-
sgn\left(Q_k(t)-Q_j(t)\right)\Big]
     e^{-|Q_k(t)-Q_j(t)|}. \label{td1}
\eeq
On the other hand, from the Neumann-like system $(F_{-2})_D$ 
we have
\beq
& & \dot{q}_j(x,t)=\{q_j, F_{-2}\}_D \nonumber\\
&=& \la_j^{-1}p_j-\left<q,p\right>q_j \nonumber\\
&=&
\frac{1}{2}q_j\Big(\sum_{k=1}^NP_k(t)sgn(x-Q_k(t))e^{-|x-Q_k(t)|}-\la_j^{-1}sgn(x-Q_j(t))\Big). \label{td2}
\eeq
Apparently, (\ref{td1}) = (\ref{td2}) iff when $x=Q_j(x,t) $, i.e. for
other $x$, they do not equal. Thus, the peakon system (\ref{PQ1}) is
not the Neumann-like Hamiltonian system $(F_{-2})_D$, either.

So, by the above analysis, we conclude: {\bf the two solutions (\ref{chu1}) and (\ref{chu2})
of the CH equation (\ref{6.4.39}) are not gauge equivalent}.
In the next subsection we will concretely solve Eq. (\ref{chu2})
on $\M$ in the form of Riemann-Theta functions.

\subsection{Algebro-geometric solution of the  CH equation on $\M$}
Now, let us re-consider the Hamiltonian system $(H_{-1})_D$ on $\M$
under the substitution of $\la \rightarrow \la^{-1}, 
\la_j \rightarrow \la^{-1}_j$ (here we choose non-zero $\la, \la_j$).
Then, the Lax matrix (\ref{GL-}) has the following simple form:
\beq
L_-(\la)=-\la^3L_{\rm CH}(\la),
\eeq
where 
\beq
L_{\rm CH}(\la)&=&\ju{0}{-\la^{-1}}{-\frac{1}{4}\la^{-1}}{0}+
\sum_{j=1}^{N}
\frac{\la^{-1}}{\la-\la_j}
              \ju{p_jq_j}{-q_j^2}{p_j^2}{p_jq_j}\nonumber\\
           &\equiv &\ju{A_{\rm CH}(\la)}{B_{\rm CH}(\la)}
{C_{\rm CH}(\la)}{-A_{\rm CH}(\la)},
 \label{LaxCH}
\eeq
and  the symplectic submanifold $\M$ becomes
\beq 
M=\left\{(q,p)\in \R^{2N}\left|\ F\equiv \frac{1}{2}\left(
\left<\La^{-1} q,q\right>-1\right)=0,\right.
G\equiv \left<\La^{-1} q,p\right>=0\right\},\nonumber\\
 \label{Symsub}
\eeq
where $\La^{-1}=diag(\la^{-1}_1,...,\la_N^{-1})$.

A direct calculation
yields to the following theorem.
 
\begin{theorem}
On the symplectic submanifold $M$ the Hamiltonian system $(H_{-1})_D$ 
defined by Eq. (\ref{H-1D}) has the Lax representation:
\beq
\frac{\pa }{\pa x}L_{\rm CH}=[M_{\rm CH},L_{\rm CH}],
\eeq
where
\beq
M_{\rm CH}=\ju{0}{1}{\frac{1}{4}
-\frac{1+4\left<\La p,p\right>}{4\left<\La^2 q,q \right>}\la^{-1}
}{0}.
\eeq
\label{Lax-theorem}
\end{theorem}

{\bf Notice}: On $M$ the Hamiltonian system $(H_{-1})_D$
is $2N-2$-dimensional, that is, only there exist $2N-2$
independent dynamical variables in all $2N$ variables
$q_1,...,q_N;\ p_1,...p_N$. Without loss of generality,
we assume the Hamiltonian system $(H_{-1})_D$ has the 
independent dynamical variables $q_1,...,q_{N-1};\ p_1,...,p_{N-1}\ (N>1)$ on $M$. Then, on $M$ we have
\beq
q^2_N&=&\la_N-\sum_{j=1}^{N-1}\frac{\la_N}{\la_j}q_j^2, \label{q_N}\\
p_N&=&-\sum_{j=1}^{N-1}\frac{\la_N}{\la_j q_N}q_jp_j, \label{p_N}
\eeq
where  $q_N$ in the latter is given by the former in terms of 
$q_1,...,q_{N-1}$. 

We will concretely give the expression
$u=\left<q(x,t),q(x,t)\right>, \ t=t_{-2}$ in an explicit
form.
By Eq. (\ref{q_N}),
We rewrite the entry  $B_{\rm CH}\left(\la\right)$ in the Lax matrix
(\ref{LaxCH}) as
\beq
B_{\rm CH}\left(\lambda\right)&=&-\frac{1}{\la-\la_N}
  \left(1+\sum_{j=1}^{N-1} \frac{(\la_j-\la_N)\la^{-1}_j}
       {\la-\la_j}q_j^2\right)\nonumber\\
&\equiv& 
-\frac{ Q\left(\lambda\right)}{K\left(\lambda\right)},
\eeq
where
\beq
Q(\la)&=&
\prod_{j=1}^{N-1}\left(\lambda-\lambda_{j}\right)
+\sum_{j=1}^{N-1} \frac{\la_j-\la_N}{\la_j}q^2_j
\prod_{k=1,k\not=j}^{N-1}\left(\lambda-\lambda_{k}\right), \label{Qq1}\\
K\left(\lambda\right)&=&
\prod_{\alpha=1}^{N}\left(\lambda-\lambda_{\alpha}\right).
\eeq

Apparently, $Q(\la)$ is a $N-1\ (N>1)$ order polynomial of $\la$.
Choosing its $N-1$ distinct  zero points $\mu_1,\ldots,\mu_{N-1}$, we have
\begin{eqnarray}
Q\left(\lambda\right)&=&\prod_{j=1}^{N-1}\left(\lambda-\mu_{j}\right), \label{Qq2}\\
\left<q,q\right>&=&\sum_{\alpha=1}^{N}\lambda_{\alpha}-\sum_{j=1}^{N-1}\mu_j. 
\label{Qq}
\end{eqnarray}
Additionally, choosing $\la=\la_j$ in Eqs. (\ref{Qq1}) and (\ref{Qq2})
leads to an explicit form of $q_j$ in terms of $\mu_l $:
\beq
q_j^2=\al_{jN}\prod_{l=1}^{N-1}(\la_j-\mu_l), \ \
\al_{jN}=\frac{\la_j}{\prod_{k=1}^{N}(\la_j-\la_k)}, \label{qj1}
\eeq
which is similar to the result in Ref. \cite{AC}. By Eq. (\ref{Qq}),
we get an identity about $\mu_l$:
\beqq
\sum_{j=1}^N\al_{jN}\prod_{l=1}^{N-1}(\la_j-\mu_l)=
\sum_{\alpha=1}^{N}\lambda_{\alpha}-\sum_{j=1}^{N-1}\mu_j. 
\eeqq

\begin{remark}
The dynamical variable $p_j$ corresponding to $q_j$ is
\beq
p_j=\frac{d q_j}{d x}
=- \frac{\al_{jN}}{2q_j}\sum_{k=1}^{N-1}\frac{d \mu_k}{d x}
\prod_{l=1,l\not=k}^{N-1}(\la_j-\mu_l),
\eeq
therefore,
\beq
 p_j^2=\frac{\al_{jN}}{4\prod_{l=1}^{N-1}(\la_j-\mu_l)}
\Big(\sum_{k=1}^{N-1}\frac{d \mu_k}{d x}
\prod_{l=1,l\not=k}^{N-1}(\la_j-\mu_l) \Big)^2. \label{pj1}
\eeq
Substituting Eqs. (\ref{Qq}) and (\ref{pj1}) into the Hamiltonian
$
H_{-1}= \frac{1}{2}\left<p,p\right>-\frac{1}{8}\left<q,q\right>$
directly gives an expression in terms of $\mu_l$:
\beq
H_{-1}=\frac{1}{8}\sum_{j=1}^N\frac{\al_{jN}}{\prod_{l=1}^{N-1}(\la_j-\mu_l)}
\Big(\sum_{k=1}^{N-1}\frac{d \mu_k}{d x}
\prod_{l=1,l\not=k}^{N-1}(\la_j-\mu_l) \Big)^2+\frac{1}{8}
\sum_{j=1}^{N-1}\mu_j
-\frac{1}{8}\sum_{k=1}^{N}\lambda_{k}.\nonumber \\
\eeq
{\sf This is evidently different  from the Hamiltonian function
in Ref. \cite{AC}} (see there section 3). Here our $H_{-1}$ 
comes from the nonlinearized CH spectral problemn, i.e. it is
composing of a Neumann-like system on $M$, which is shown  integrable
in subsection 3.3 by $r$-matrix process. It is because of this
difference that our parametric solution does not include
the peakons (also see last subsection) and  we will in the following
procedure present 
a class of new algebro-geometric solution  for the CH equation
constrained on the symplectic
submanifold $M$. 
\end{remark}

Let
\be
\pi_j=A_{\rm CH}\left(\mu_j\right), \ \ j=1,...,N-1,
\ee
then it is easy to prove the following proposition.

    \begin{proposition} \quad 
\begin{equation}
\{\mu_{i}, \mu_{j}\}_D=\{\pi_{i}, \pi_{j}\}_D=0,\quad  
\{\pi_{j}, \mu_{i}\}_D=\delta_{ij},
 \quad  i, j=1, 2, ..., N-1,
\end{equation}
i.e. $\pi_j,\mu_j$ are conjugated, and thus they
are the variables which can be separated {\rm \cite{SK}}.
\end{proposition}

Write
\beq
-\det L_{\rm CH}\left(\la\right)
& =& A_{\rm CH}^2\left(\la\right)+B_{\rm CH}\left(\la\right)C_{\rm CH}
\left(\la\right) \nonumber\\
& =& \frac{1}{\la^{2}}\left(\frac{1}{4}+\sum^N_{\alpha=1}\frac{E_{\alpha}}
{\la-\la_{\alpha}}\right)\nonumber \\
&=&\frac{1}{\la(\la-\la_N)}
  \left(\frac{1}{4}+\sum_{\al=1}^{N-1} \frac{(\la_{\al}-\la_N)\la^{-1}_{\al}}
       {\la-\la_{\al}}E_{\al}\right)\nonumber\\
& \equiv& \frac{ P\left(\lambda\right)}{\la K\left(\lambda\right)}, \label{Lch}
\eeq
where $E_{\alpha}$ is defined by 
\beq
E_{\alpha}&=&-p_{\al}^2+\frac{1}{4}q_{\al}^2-\Gamma_{\al},\ \   
 \Gamma_{\al}=\sum^N_{k=1,k\not=\al}
\frac{(p_{\al}q_k-q_{\al}p_k)^2}{\la_{\al}-\la_k}, \label{Ech}\\
P(\la)&=&\frac{1}{4}
\prod_{j=1}^{N-1}\left(\lambda-\lambda_{j}\right)
+\sum_{j=1}^{N-1} \frac{\la_j-\la_N}{\la_j}q^2_j
\prod_{k=1,k\not=j}^{N-1}\left(\lambda-\lambda_{k}\right), \label{Pch}
\eeq
and obviously $P\left(\lambda\right)$ is an $N-1$ order polynomial of $\lambda$ whose
first term's coefficient
is $\frac{1}{4}$.
Then we have 
\beq 
\pi_j^2=\frac{ P\left(\mu_j\right)}{\mu_j K\left(\mu_j\right)},\
j=1,\ldots,N-1. \label{Pij}
\eeq

{\bf Notice}: On $M$ we always have
$\sum_{\al=1}^N\la^{-1}_{\al}E_{\al}=\frac{1}{4}$. Therefore, we
assume $E_1,...,E_{N-1}$ are independent. Then, 
$$E_N=\frac{1}{4}\la_N-\sum_{\al=1}^{N-1}\frac{\la_N}{\la_{\al}}E_{\al}.
$$
Actually in Eq. (\ref{Lch}) we used this fact.

Now, we choose the generating function
\beq
W&=&\sum_{j=1}^{N-1}W_j\left(\mu_j,\{E_{\alpha}\}_{\alpha=1}^{N-1}\right)
\nonumber\\
&=&\sum_{j=1}^{N-1}
\int_{\mu_0}^{\mu_j}\sqrt{\frac{ P\left(\lambda\right)}
{\la K\left(\lambda\right)}} {\rm d} \lambda,
\eeq
where $\mu_0$ is an arbitrarily given constant.
Let us view $E_{\alpha} \ \left(\alpha=1,\ldots,N-1\right)$ 
as actional variables,
then angle-coordinates $Q_{\alpha}$ are chosen as
$$
 Q_{\alpha}=\frac{\partial W}{\partial E_{\alpha}}, \quad \alpha=1,\ldots,N-1,
$$
i.e.
\beq
Q_{\alpha}&=& \sum_{k=1}^{N-1} \int_{\mu_0}^{\mu_k}
\frac{K(\la)}{2
\sqrt{\la K\left(\lambda\right)P\left(\lambda\right)}}
\cdot\frac{\la_{\al}-\la_N}
{\la_{\al}(\la-\la_{\al})(\la-\la_N)} {\rm d}\la \nonumber\\
&\equiv& \frac{\la_{\al}-\la_N}
{\la_{\al}}
 \sum_{k=1}^{N-1} \int_{\mu_0}^{\mu_k}\tilde{\omega}_{\alpha},
\eeq
where
\beq
\tilde{\omega}_{\alpha}
=
\frac{ \prod^{N-1}_{k\not={\alpha},k=1}\left(\lambda-\lambda_k
\right)}{2\sqrt{\la K\left(\lambda\right)P\left(\lambda\right)}} {\rm
d}\lambda,\ \ 
 \alpha=1,\ldots,N-1.\nonumber
\eeq

    Therefore, on the symplectic submanifold $\left(M^{2N-2},dE_{\alpha}\wedge dQ_{\alpha}\right)$
the Hamiltonian function 
\beq
H_{-1}&=&\frac{1}{2}\left<p,p\right>-\frac{1}{8}\left<q,q\right>=-\frac{1}{2}
\sum_{\alpha=1}^NE_{\alpha} \nonumber\\
&=&-\frac{1}{8}\la_N-\frac{1}{2}\sum_{\al=1}
^{N-1}\frac{\la_{\al}-\la_N}{\la_{\al}}E_{\al}
\eeq
 produces a
linearized $x$-flow of the CH equation
\be
\left\{\begin{array}{l}
Q_{\alpha,x}=\frac{\partial H_{-1}}{\partial E_{\alpha}}
   =-\frac{\la_{\al}-\la_N}{2\la_{\al}},\\
E_{\alpha,x}=0,
\end{array}\right.
\ee
as well as
the Hamiltonian function 
\beq
F_{-2}&=&\frac{1}{2}\left<\La p,p\right>-\frac{1}{8}\left<\La
  q,q\right>+\frac{1}{2}\left(\left<q,q\right>\left<p,p\right>
-\left<q,p\right>^2\right) = -\frac{1}{2}
\sum_{\alpha=1}^N\la_{\al}E_{\alpha} \nonumber\\
&=&-\frac{1}{8}\la_N^2-\frac{1}{2}\sum_{\al=1}
^{N-1}\frac{\la_{\al}^2-\la_N^2}{\la_{\al}}E_{\al}
\eeq
 yields a
linearized $t$-flow of the CH equation
\be
\left\{\begin{array}{l}
Q_{\alpha,t}=\frac{\partial F_{-2}}{\partial E_{\alpha}}
   =-\frac{\la_{\al}^2-\la_N^2}{2\la_{\al}},\\
E_{\alpha,t}=0.
\end{array}\right.
\ee

The above two flows imply
\beq
Q_{\al}&=&\frac{\la_N-\la_{\al}}{2\la_{\al}}\Big[
       x+(\la_N+\la_{\al})t-2Q^0_{\al}\Big], \label{Qal}\\
E_{\al}&=& {\rm constant,} \ \ \al=1,...,N-1,
\eeq
where $Q_{\alpha}^0$ is an arbitrarily chosen  constant. Therefore we
have
\beq
 \sum_{k=1}^{N-1} \int_{\mu_0}^{\mu_k}\tilde{\omega}_{\alpha}&=&
-\frac{1}{2}\Big[
       x+(\la_N+\la_{\al})t\Big]+Q^0_{\al}.
\eeq

    Choose a basic system of closed paths $\alpha_i,\beta_i \ \left(i=1,\ldots,N-1\right)$
of Riemann surface $\bar{\Gamma}$:\ $\mu^2=\la P\left(\lambda\right)K\left(\lambda\right)$ with $N-1$
handles. $\tilde{\omega}_j$ ($j=1,\ldots,N-1$) are exactly $N-1$ linearly
independent holomorphic differentials
 of the first kind
on the Riemann surface $\bar{\Gamma}$. Let
$\tilde{\omega}_j$ be normalized as $\omega_j=\sum_{l=1}^{N-1}
r_{j,l}\tilde{\omega}_l$,
i.e. $\omega_j$ satisfy
$$ \oint_{\alpha_i}\omega_j=\delta_{ij},\quad  \oint_{\beta_i}\omega_j=B_{ij}, $$
where $B=\left(B_{ij}\right)_{(N-1)\times (N-1)}$ is symmetric and the imaginary part ${\rm Im} B$ of $B$
is a positive definite matrix.

    By Riemann Theorem \cite{GH} we know:
    $\mu_k$ satisfy
 \beqq 
\sum_{k=1}^{N-1} \int
_{\mu_0}^{\mu_k}\omega_j&=&\phi_j, \nonumber\\
 \phi_j&=&\phi_j\left(x,t\right)\stackrel{\rm d}{=}
\sum_{l=1}^{N-1} r_{j,l}
\left(Q^0_{l}-\frac{1}{2}\Big[
       x+(\la_N+\la_{l})t\Big]\right),  \\ 
 & & \ \ \ \ \ j=1,\ldots,N-1, \nonumber 
\eeqq
 iff
$\mu_k$ are the zero points of the Riemann-Theta function
   $\tilde{\Theta}\left(P\right)=\Theta\left(A\left(P\right)- \phi-\K\right)$
which has exactly $N-1$ zero points, where 
\beqq
A\left(P\right)&=&
\left(\int^P_{P_0}\omega_1,\cdots,\int^P_{P_0}\omega_{N-1}\right)^T,\\
\phi&=&\phi\left(x,t\right)
=\left(\phi_1\left(x,t\right),\cdots,\phi_{N-1}\left(x,t\right)\right)^T,\eeqq
$\K=(K_1,...,K_{N-1})^T\in \C^{N-1}$ 
is the Riemann constant vector,
$P_0$ is an arbitrarily given point on Riemann
surface $\bar{\Gamma}$ ($\Theta$-function and
the related properties are seen in Appendix).

    Because of \cite{D}
\be
\frac{1}{2\pi i}\oint_\gamma\lambda d\ln\tilde{\Theta}\left(P\right)=C_1\left(\bar{\Gamma}\right),
\ee
where the constant $C_1\left(\bar{\Gamma}\right)$ has none thing to do with $\phi$;
$\gamma$ is the boundary of simple connected domain obtained through cutting
the Riemann surface $\bar{\Gamma}$ along closed paths $\alpha_i,\beta_i$. Thus,
we have a key equality
\begin{equation}
\sum_{k=1}^{N-1}\mu_k=C_1\left(\bar{\Gamma}\right)-{\rm Res} _{\lambda=\infty_1}\lambda d\ln \tilde{\Theta}\left(P\right)
-{\rm Res} _{\lambda=\infty_2}\lambda d\ln \tilde{\Theta}\left(P\right), \label{mun}
\end{equation}
where ${\rm Res} _{\la=\infty_k}$ $\left(k=1,2\right)$ mean
 the residue at points $\infty_k$:
   \beqq
& &\infty_1\stackrel{\rm d }{=}
    \left(0,\left.\sqrt{z^{-1}P\left(z^{-1}\right)K\left(z^{-1}\right)}\right|_{z=0}\right),\\
& & \infty_2
\stackrel{\rm d }{=}
\left(0,-\left.\sqrt{z^{-1}P\left(z^{-1}\right)K\left(z^{-1}\right)}\right|_{z=0}\right).
\eeqq

Now, we need to calculate the above two residues.

\begin{lemma}

\begin{eqnarray} {\rm Res} _{\lambda=
\infty_1}\lambda d\ln \tilde{\Theta}\left(P\right)&=&-2\frac{\partial }{\partial x}
                         \ln \Theta \left(\phi+\K+\eta_1\right),\\
        {\rm Res} _{\lambda=\infty_2}\lambda d\ln
                         \tilde{\Theta}\left(P\right)&=& 2\frac{\partial }{\partial x}
    \ln \Theta \left(\phi+\K+\eta_2\right),
\end{eqnarray}
where $\eta_1, \eta_2 $ are two different $N-1$ dimensional constant vectors.
\end{lemma}

    {\bf Proof }
       \ Consider the following smooth superelliptic curve
        $\bar{\Gamma}$:  $\mu^2=\la P\left(\lambda\right)K\left(\lambda\right)$.
    Because $\la P\left(\lambda\right) K\left(\lambda\right)$
    is a $2N$-th order polynomials with respect to $\lambda$,
    $\infty$ is not a branch point, i.e. on $\bar{\Gamma}$ there are
    two infinity points
$\infty_1$ and 
$ \infty_2$.
All points $P$ on $\bar{\Gamma}$ are denoted by $\left(\lambda,\pm\mu\right)$.
On $\bar{\Gamma}$ we choose a group of basis of normalized closed paths
$\alpha_1,\cdots,\alpha_{N-1};\beta_1,
\cdots,\beta_{N-1}$. They are mutually independent,
and their intersection number are
$$\alpha_i\circ\alpha_j
=\beta_i\circ\beta_j=0,\quad \alpha_i\circ\beta_j=\delta_{ij}.$$
It is easy to see that
$\tilde{\omega}_j$ $\left(j=1,...,N-1\right)$ are $N-1$ linearly independent
holomorphic differential forms on
$\bar{\Gamma}$.

    Let us now come to calculate
    the residus of $\lambda d\ln \tilde{\Theta}\left(P\right)$
    at the two infinity points: $\infty_1,\infty_2$.
At $\infty_1$, the $j$-th variable $I_j$ of $\tilde{\Theta}\left(z\right)$
produces the following result through multiplying by $-1$
(please note that the local coordinate at $\infty_1$ and
$\infty_2$ is $z=\lambda^{-1}$):
\begin{eqnarray*}
-I_j&=&\phi_{j}+K_j+\eta_{1,j}-\sum^{N-1}_{l=1}r_{j,l}\int^z_0\tilde{\omega}_l\\
      &=&\phi_{j}+K_j+\eta_{1,j}+\left.
\sum^{N-1}_{l=1}r_{j,l}
\int^z_0
         \frac{\prod^{N-1}_{\alpha\not=l,\alpha=1}
         \left(\lambda-\lambda_\alpha\right)}
               {2\sqrt{\la P\left(\lambda\right)K\left(\lambda\right)}}
               \right|_{\lambda=z^{-1}}z^{-2}{\rm d}z\\
      &=&\phi_{j}+K_j+\eta_{1,j}+
\sum^{N-1}_{l=1}r_{j,l}
\int^z_0
         \frac{z^{-N}-(\sum^{N-1}_{j=1}\la_j-\la_l)z^{-N+1}+\cdots}
           {\sqrt{z^{-2N}+\cdots}}{\rm d} z\\
      &=&\phi_{j}+K_j+\eta_{1,j}+
\sum^{N-1}_{l=1}r_{j,l}\int^z_0
         \frac{1+ O\left(z\right)}
              {\sqrt{1+O\left(z\right)}}{\rm d}z  \\
      &=&\phi_{j}+K_j+\eta_{1,j}+\sum^{N-1}_{l=1}r_{j,l}z+O\left(z^2\right),
\end{eqnarray*}
where 
$\eta_{1,j}=\int^{P_0}_{\infty_1}\omega_j,\ j=1,\cdots N-1$.

    Because
 \beqq
\frac{\partial \Theta}{\partial x}=
\frac{1}{2}\sum^{N-1}_{j=1}
    \sum^{N-1}_{l=1}\frac{\partial \Theta}{\partial I_j}r_{j,l}
\eeqq
and $\tilde{\Theta}\left(z\right)$
has the expansion formula
$$\tilde{\Theta}\left(z\right)=\Theta\left(\phi+\K+\eta_1\right)
-\sum^{N-1}_{j=1}\sum^{N-1}_{l=1}
    \frac{\partial \Theta}{\partial I_j}r_{j,l}z+O\left(z^2\right),$$
where $\eta_1=(\eta_{1,1},...,\eta_{1,N-1})^T$.
Therefore,
$$\tilde{\Theta}\left(z\right)=
\Theta\left(\phi+\K+\eta_1\right)-2\frac{\partial \Theta}{\partial x}z
     +O\left(z^2\right).$$
So, we obtain the following residue
\begin{eqnarray*}
{\rm Res} _{\lambda=\infty_1}\lambda d\ln \tilde{\Theta}\left(P\right)
&=&{\rm Res} _{z=0}z^{-1}d\ln \tilde{\Theta}\left(z\right)={\rm Res} _{z=0}\frac{1}{z}\frac{\tilde{\Theta}_z\left(z\right)}{\tilde{\Theta}\left(z\right)}\\
&=&{\rm Res} _{z=0}\frac{1}{z}\frac{-2\Theta_x+O\left(z\right)}
 {\Theta-2\Theta_xz+O\left(z^2\right)}\\
&=&{\rm Res} _{z=0}\frac{1}{z}\frac{-2\Theta_x+O\left(z\right)}
{\Theta\left(1-2\Theta^{-1}\Theta_xz+O\left(z^2\right)\right)}\\
&=&\frac{-2\Theta_x}{\Theta}=-2\frac{\partial }{\partial x}
\ln \Theta \left(\phi+\K+\eta_1\right).
\end{eqnarray*}

    In a similar way, we can obtain the second residue formula. \hfill $\rule{2mm}{4mm}$

$ $

   So, by Eq. (\ref{Qq}) and this Lemma, we immediately have
\begin{equation}
\left<q\left(x,t\right),q\left(x,t\right)\right>
=\sum_{\alpha=1}^N\lambda_{\alpha}-C_1\left(\bar{\Gamma}\right)
+2\frac{\pa}{\pa x}\left(\ln\frac
{\Theta\left(\phi+\K+\eta_2\right)}{\Theta\left(\phi
+\K+\eta_1\right)}\right),
\end{equation}
where the $j$-th component of $\eta_i \ \left(i=1,2\right)$ is
$\eta_{i,j}=\int^{P_0}_{\infty_i}\omega_j$. 

So, the  CH equation (\ref{6.4.39}) on $M$ has the following
explicit solution, called the {\bf algebro-geometric
solution}
\beq
u(x,t)=R
+2\frac{\pa}{\pa x}\left(\ln\frac
{\Theta\left(\phi+\K+\eta_2\right)}{\Theta\left(\phi
+\K+\eta_1\right)}\right), \label{CHsolution}
\eeq
where
$R=\sum_{\alpha=1}^N\lambda_{\alpha}-C_1\left(\bar{\Gamma}\right)$
is a constant.

\begin{theorem}
   The algebro-geometric solution of the {\rm CH}  equation
   {\rm (\ref{6.4.39})} on $M$ can be given through
Eq. (\ref{CHsolution}).
\end{theorem}

\begin{remark}
Here the algebro-geometric solution (\ref{CHsolution})
is smooth and 
occurs in the $x$-direction (spacial variable) derivative, and
apparently differs from the piecewise smooth 
algebro-geometric solution
in the $t$-direction derivatives given in Ref. \cite{AC}.
It is also different from the results in Ref. \cite{AC1,AF1,AF2,GH1},
becasue we are studying the CH equation constrained on $M$.
In the paper, we do not need to calculate each $q_j \ (j=1,...,N-1)$
but the sum $\sum^{N-1}_{j=1}q^2_j$, which we know
from  Eqs. (\ref{Qq}) and (\ref{u2-}). 
From the above subsection's comparison and these comments, 
therefore we
think Eq. (\ref{CHsolution})
is a class of new solution of the  CH equation
(\ref{6.4.39}) on $M$. Apparently, Eq. (\ref{CHsolution})
is more simpler in the form than in Ref. \cite{AC}.
\end{remark}

\section{Positive order CH hierarchy,  integrable Bargmann system, and
  parametric solution}
\setcounter{equation}{0}

\subsection{Positive order CH hierarchy}

Let us now give the positive order hierarchy of Eq. (\ref{CHL1})
through employing the kernel element of the Lenard's operator
 $J$. 

Obviously,
$G_0=cm^{-\frac{1}{2}}$ form all  kernal elements of $J$,
where $c=c(t_n)\in C^{\infty}(\R)$
is an arbitrarily given fuctions with respect to
the time variables $t_n \ (n\ge 0, n\in \Z)$, but independent of the spacial
variable $x$.  
$G_0$ 
and the recursion operator (\ref{6.4.4}) yield
 the following  hierarchy of the CH spectral problem (\ref{CHL1})
\be
m_{t_k}
=cJ{\cal L}^k\cdot m^{-\frac{1}{2}}, \ \ k=0,1,2,\ldots, \label{6.4.19}
\ee
where the operators $J$ and ${\cal L}$ are defined by Eqs. (\ref{6.4.3})
 and (\ref{6.4.4}), respectively, and ${\cal L}$ and $J$ have a further
product form
\beq
{\cal L}&=&J^{-1}K=\frac{1}{2}m^{-\frac{1}{2}}\pa^{-1} m^{-\frac{1}{2}}
(\pa-\pa^3)\\
J^{-1}&=&\frac{1}{2}m^{-\frac{1}{2}}\pa^{-1} m^{-\frac{1}{2}}
\eeq
Because Eq. (\ref{6.4.19}) is related to the Camassa-Holm spectral problem
(\ref{CHL1}) for the case of $k\ge0,\ k\in \Z$, it is  called 
{\bf the positive order Camassa-Holm (CH) hierarchy}.
Eq. (\ref{6.4.19}) has the following representative equations
\beq
m_{t_0}&=&0,  \ \ {\rm trivial \ case}, \label{6.4.20} \\
m_{t_1}&=&-c \left(m^{-\frac{1}{2}}\right)_{xxx}+
c\left(m^{-\frac{1}{2}}\right)_x. \label{6.4.21}
\eeq

Apparently, with $c=-1$ Eq. (\ref{6.4.21}) becomes 
a Dym type equation
\beq
m_{t_1}&=& \left(m^{-\frac{1}{2}}\right)_{xxx}-
\left(m^{-\frac{1}{2}}\right)_x, \label{6.4.21k}
\eeq
which has more an extra term $-
\left(m^{-\frac{1}{2}}\right)_x $ than the usual Harry-Dym
equation $m_{t_1}= \left(m^{-\frac{1}{2}}\right)_{xxx}$.
Therefore, Eq. (\ref{6.4.19})
gives an extended  Dym
hierarchy corresponding to the isospectral case: $\la_{t_k}=0$.

By Theorem \ref{Th12}, the whole positive order CH  hierarchy (\ref{6.4.19}) has
the zero curvature representation
\beq
U_{t_k}-V_{k,x}+[U,V_k]&=&0,\ k>0, \  k\in \Z \label{NUV1}\\
V_k&=&\sum_{j=0}^{k-1} V_j\la^{k-j-1},
\eeq
where $U$ is given by Eq. (\ref{CHL1}), and  
$V_j=V(G_j)$ is given by Eq. (\ref{6.4.15}) with 
$G=G_j={\cal L}^{j}\cdot G_{0}=c{\cal L}^j\cdot m^{-\frac{1}{2}}, 
\ j>0,\ j\in \Z$.
Thus, all nonlinear equations in the positive CH hierarchy (\ref{6.4.19}) are
integrable. Therefore we obtain the following theorem.

\begin{theorem}
The positive order CH hierarchy (\ref{6.4.19})
 possesses the Lax pair
\beq 
\left\{\begin{array}{l}
y_{x}=\ju{0}{1}{\frac{1}{4}-\frac{1}{2}m\la}{0}y,\\
y_{t_k}=\sum_{j=0}^{k-1}
\ju{-\frac{1}{2}G_{j,x}}{-G_j}{\frac{1}{2}G_{j,xx}-\frac{1}{4}G_j+\frac{1}{
2}m\la G_j}{\frac{1}{2}G_{j,x}}\la^{k-j}y \\
\ \ \ \ \ \ \ \ \ \  k=0,1,2,....
\end{array}\right. \label{Laxch1}
\eeq
\label{Th2}
\end{theorem}

 Eq. (\ref{Laxch1}) can be reduced to the following Lax pair
\beq 
\left\{\begin{array}{l}
\psi_{xx}=\frac{1}{4}\psi-\frac{1}{2}m\la\psi\\
\psi_{t_{k}}=\sum_{j=0}^{k-1}
\left(\frac{1}{2}G_{j,x}\la^{k-j}\psi-G_j\la^{k-j}\psi_x\right), \ k=0,1,...,
\end{array}\right.
\eeq

In particular, the Dym type  equation (\ref{6.4.21k}) has the Lax pair
\beq 
\left\{\begin{array}{l}
\psi_{xx}=\frac{1}{4}\psi-\frac{1}{2}m\la\psi\\
\psi_{t_{1}}=-\frac{1}{2} \left(m^{-\frac{1}{2}}\right)_x
\la\psi+ m^{-\frac{1}{2}}\la\psi_x.
\end{array}\right.
\eeq

\begin{remark}
This Lax pair coincides with 
the one by the usual method
of finite power expansion with respect to spectral parameter
$\la$.
However, we here present the positive hierarchy (\ref{6.4.19}) mainly by
 the Lenard's operators pair satisfying Eq. (\ref{6.4.2}). Due to containing
the spectral gradient  $\nabla\la$ in Eq. (\ref{6.4.2}),
this procedure of generating evolution equations from a given spectral problem
is called  the spectral gradient method.
\end{remark}
In this method, how to determine
a pair of Lenard's operators associated with the given spectral problem
mainly depends on the concrete forms of spectral problems and spectral
gradients, and some computational techniques.
From this method, we have already derived the negative order
CH hierarchy and  the positive order CH hierarchy.

\subsection{$r$-matrix structure and integrability
for the Bargmann CH system}

Consider the following matrix  (called `positive' Lax matrix)
\begin{equation}
L_+\left(\lambda\right)=\left(\begin{array}{lr}
A_+\left(\lambda\right) & B_+ \left(\lambda\right)\\
C_+\left(\lambda\right) & -A_+\left(\lambda\right)
\end{array}\right)  \label{GL}
\end{equation}
where
\begin{eqnarray}
A_+\left(\lambda\right)&=&
           \sum_{j=1}^{N}\frac{\la_jp_{j}q_{j}}{\la-\la_{j}}, \\
B_+\left(\lambda\right)&=&-\sum_{j=1}^{N}\frac{\la_jq_{j}^{2}}{\la-\la_{j}},\\
C_+\left(\lambda\right)&=&\frac{1}{2\left<\La q,q\right>}+
\sum_{j=1}^{N}\frac{\la_jp_{j}^{2}}{\la-\la_{j}}.
\end{eqnarray}

We calculate the determinant of $L_+(\la)$:

\begin{eqnarray}
-\frac{1}{2}\det L_+(\la)&=&
\frac{1}{4}{\rm Tr}L_+^{2}(\la)
 =\frac{1}{2}\left(A_+^2\left(\lambda\right)
 +B_+\left(\lambda\right)C_+\left(\lambda\right)\right) \nonumber\\
                &=& \sum_{j=1}^{N}\frac{E_{j}^+}{\la-\la_{j}}
\label{detL}
\end{eqnarray}
where ${\rm Tr}$ stands for the trace of a matrix, and {\normalsize
\begin{eqnarray}
E_j^+&=& -\frac{1}{4\left<\La q,q\right>}\la_jq_j^2
       -\frac{1}{2}\Gamma_j^+, \ \ \ j=1,2,...,N,\\
   & & \Gamma_j^+ =\sum_{l\not=j,l=1}^N\frac{\la_j\la_l(p_jq_l-p_lq_j)^2}
   {\la_j-\la_l}.
\end{eqnarray}}

Let
\beq
F_k=\sum_{j=1}^{N}\la^{k}_jE_j^+,\ \ k=0,1,2,\ldots,
\eeq
then it reads
\beq
F_{k}&=&-\frac{\left<\La^{k+1}q,q\right>}
         {4\left<\La q,q\right>}+\frac{1}{2}\sum_{j=0}^{k-1}\left(
     \left<\La^{j+1}q,p\right>\left<\La^{k-j}q,p\right>-
     \left<\La^{j+1}q,q\right>\left<\La^{k-j}p,p\right>\right),\nonumber\\
                          \label{Fk+}
\eeq
where $k=0,1,2....$
A long but direct computation leads to
the following key equalities:
\beqq
\{A_+(\la),A_+(\mu)\}&=&\{B_+(\la),B_+(\mu)\}=0,\\
\{C_+(\la),C_+(\mu)\}&=& \frac{2}{\left<\La q,q\right>^2}
            \left(\la A_+(\la)- \mu A_+(\mu)\right),\\
\{A_+(\la),B_+(\mu)\}&=&
          \frac{2}{\mu-\la}(\mu B_+(\mu)-\la B_+(\la)),\\
\{A_+(\la),C_+(\mu)\}&=&
 \frac{2}{\mu-\la}(\la C_+(\la)-\mu C_+(\mu))
             -\frac{1}{\left<\La q,q\right>^2}\la B_+(\la),\\
\{B_+(\la),C_+(\mu)\}&=&
         \frac{4}{\mu-\la}(\mu A_+(\mu)-\la A_+(\la)).
\eeqq

$ $

Let $L_{1}^+\left(\la\right)=L_+\left(\lambda\right)\otimes I_{2\times2}$,
$L_{2}^+\left(\mu\right)=I_{2\times2}\otimes L_+\left(\mu\right)$,
where $L_+\left(\la\right),\ L_+\left(\mu\right)$ are given through
Eq. (\ref{GL}).
In the following, we search for a $4\times 4$   $r$-matrix
structure
$r_{12}^+\left(\lambda,\mu\right)$ such that the fundamental Poisson bracket
\cite{FT}:
\begin{equation}
\left\{L_+\left(\lambda\right) \stackrel{\otimes}{,}
L_+\left(\mu\right)\right\}=\left[r_{12}^+\left(\lambda,\mu\right),
L_{1}^+\left(\lambda\right)\right]-\left[r_{21}^+\left(\mu,\lambda\right), L_{2}^+\left(\mu\right)\right]  \label{fpb}
\end{equation}
holds, where the entries of the $4\times4$ matrix
$\left\{L_+\left(\lambda\right) \stackrel{\otimes}{,}
L_+\left(\mu\right)\right\}$ are
$$\left\{L_+\left(\lambda\right) \stackrel{\otimes}{,}
L_+\left(\mu\right)\right\}_{kl,mn}=\left\{L_+\left(\la\right)_{km},
L_+\left(\mu\right)_{ln}\right\}, \ \ k,l,m,n=1,2,$$
and
$r_{21}\left(\lambda,\mu\right)=Pr_{12}\left(\lambda,\mu\right)P,$ with
\beq
P&=&\frac{1}{2}\left(I_{2\times2}+\sum_{j=1}^3\sigma_j\otimes\sigma_j\right)=
 \left(\begin{array}{cccc}
1& 0 & 0 & 0\\
0& 0 & 1 & 0\\
0& 1 & 0 & 0\\
0& 0 & 0 & 1
\end{array}
\right), \nonumber
\eeq
where $\sigma_j'$ are the Pauli matrices.

$ $

   \begin{theorem}
\be
r_{12}^+\left(\lambda,\mu\right)=\frac{2\la}{\mu(\mu-\la)}P+
\frac{\la}{\left<\La q,q\right>^2}S^+ \label{r+}
\ee
is an   $r$-matrix   structure satisfying Eq.  (\ref{fpb}), where
\begin{eqnarray*}
S^+=\sigma_-\otimes\sigma_-= \ju{0}{0}{1}{0} \otimes \ju{0}{0}{1}{0}.
\end{eqnarray*}
\end{theorem}

$ $

In fact, the $r$-matrix satisfying Eq. (\ref{fpb})
is not unique \cite{Q1}.
Obviously, this $r$-matrix structure is also of $4\times 4$ and 
 different from the one in Ref. \cite{RB}.

%\subsection{Integrability}
%\setcounter{equation}{0}

Because there is an   $r$-matrix   structure
satisfying Eq.  (\ref{fpb}),
\beq
\left\{L^2_+\left(\la\right) \stackrel{\otimes}{,} L^{2}_+\left(\mu\right)
\right\}=\left[\bar{r}_{12}^+\left(\la,\mu\right), L_{1}^+
\left(\la\right)\right]-\left[\bar{r}_{21}^+\left(\mu,\la\right),
L_{2}^+\left(\mu\right)\right]
\eeq
where
\beqq
\bar{r}_{ij}^+\left(\lambda,\mu\right)=\sum_{k=0}^{1}\sum_{l=0}^{1}
(L_{1}^+)^{1-k}\left(\lambda\right)
(L_{2}^+)^{1-l}\left(\mu\right) r_{ij}\left(\lambda,\mu\right)
(L_{1}^+)^{k}\left(\lambda\right)
(L_{2}^+)^{l}\left(\mu\right), \\  ij=12, 21.
\eeqq
Thus,
\begin{equation}
4\left\{{\rm Tr}L^{2}_+\left(\lambda\right), {\rm Tr}L^{2}_+
\left(\mu\right)\right\}={\rm Tr}\left\{L_+^{2}
\left(\lambda\right) \stackrel{\otimes}{,}
L_+^{2}\left(\mu\right)\right\}=0.
\end{equation}
So, by Eq.  (\ref{detL}) we immediately obtain the following theorem.

\begin{theorem}
The following equalities
\beq
& &\{E_{i}^+, E_{j}^+\}=0, \  \  \{F_k, E_{j}^+\}=0, \\
& & i, j=1,2,\ldots,N,\ \ k=0,1,2,\dots, \nonumber
\eeq
hold. Hence, all Hamiltonian systems $\left(F_k\right)$
\beq
\left(F_{k}\right) :  \quad q_{t_{k}}=\{q, F_{k}\}=
\frac{\pa F_k}{\pa p}, \
\   p_{t_{k}}=\{p,F_{k}\}=-\frac{\pa F_k}{\pa q},\\
   k=0,1,2,\ldots,\nonumber
\eeq
are completely integrable.
\end{theorem}

$ $

Furthermore, we find the following Hamiltonian function
\beq
H^+=\frac{1}{2}\left<p,p\right>-\frac{1}{8}\left<q,q\right>
    -\frac{1}{4\left<\La q,q\right>} \label{H+}
\eeq
is involutive with $E_j^+, \ F_k$, i.e.
\beq
& &\{H^+, E_{j}^+\}=0, \  \  \{H^+, F_k\}=0,\\
& & \ \ \ \ \  j=1,2,\ldots,N,\ \ k=0,1,2,\dots. \nonumber
\eeq
Here, $E_j^+$ are $N$ independent functions.

Therefore,  we obtain the following results.
\begin{corollary}
The canonical Hamiltonian system $(H^+)$:
\beq
(H^+): \ \
\left\{\begin{array}{l}
q_x=\frac{\pa H^+}{\pa p}=p,\\
p_x=-\frac{\pa H^+}{\pa q}=\frac{1}{4}q-\frac{1}{2\left<\La q,q\right>^2}\La q.
\end{array}\right. \label{H++}
\eeq
is completely integrable.
\end{corollary}

    \begin{corollary} 
 All composition functions $f\left(H^+, F_k^+\right)$, $f\in
C^{\infty}\left(\R\right)$, $k=0,1,2,...$, are completely
integrable Hamiltonians.
\end{corollary}

Let
\beq
m&=& \frac{1}{\left<\La q,q \right>^2},\label{mch1}\\
\psi&=&q_j, \ \ \ \la=\la_j, \ \ j=1,...,N.
\eeq
Then, the integrable flow $(H^+)$ defined by Eq. (\ref{H++})
also exactly becomes the CH spectral problem (\ref{CH1})
with the potential function $m$.

\begin{remark}
Eq. (\ref{mch1}) is a Bargmann  constraint in the whole space
$\R^{2N}$. Therefore, {\bf the Hamiltonian system $(H^+)$ is of integrable
Bargmann type}.
\end{remark}

\subsection{Parametric solution of the positive order CH hierarchy}

In the following, we shall consider the relation between
constraint and nonlinear equations in the positive order
CH hierarchy (\ref{6.4.19}). Let us start from the following
setting
\beq
 G_0 & = &- \sum_{j=1}^N\nabla\la_j,
  \label{4.10}
\eeq
where $G_{0}=-m^{-\frac{1}{2}}$, and $ \nabla\la_j=\la_jq^2_j$
is the functional gradient of the CH spectral problem
(\ref{CHL1}) corresponding to the spectral parameter $\la_j$
($j=1,...,N$).

  Apparently Eq. (\ref{4.10}) reads
\beq
m=\frac{1}{\left<\La q,q\right>^2} \label{qm1}
\eeq
which coincides with the constraint relation (\ref{mch1}).

Since the Hamiltonian flows $(H^+)$ and $(F_k)$ are completely
integrable and their Poisson brackets $\{H^+,F_k\}=0$ ($k=0,1,2,...$),
their phase flows $g^x_{H^+},\ g^{t_k}_{F_k}$ commute \cite{AV}. Thus,
we can define their compatible solution as follows:
\beq
\left(\begin{array}{l}
q(x,t_k)\\
p(x,t_k)
\end{array}
\right)=g^x_{H^+} g^{t_k}_{F_k}\left(\begin{array}{l}
q(x^0,t_k^0)\\
p(x^0,t_k^0)
\end{array}
\right), \ \ k=0,1,2,...,
\eeq
where $x^0, \ t_k^0$ are the initial values of phase flows
$g^x_{H^+},\ g^{t_k}_{F_k}$.

\begin{theorem}
Let $q(x,t_k), \ p(x,t_k)$ be a solution of the compatible Hamiltonian systems
$(H^+)$ and $(F_k)$. Then
\beq
m=\frac{1}
{\left<\La q(x,t_k),q(x,t_k)\right>^2} \label{mr}
\eeq
satisfies the positive order CH equation
   \beq
m_{t_k} & = &-J{\cal L}^k\cdot m^{-\frac{1}{2}},
   \ \  k=0,1,2,...,
   \label{umkdv}
   \eeq
where the operators $J$ and ${\cal L}$ are given by Eqs. (\ref{6.4.3})
and (\ref{6.4.4}), respectively.
   \label{th4}
\end{theorem}
{\bf Proof}: This proof is similar to the negative case.

%\end{document}

$ $

In particular, with $k=1$, we obtain the following corollary.
\begin{corollary}
Let $q(x,t_{1}), \ p(x,t_{1})$ be a
solution of the compatible integrable Hamiltonian systems
$(H^+)$ and $(F_{1})$. Then
\beq
m=m(x,t_{1})=\frac{1}{\left<\La q(x,t_{1}),q(x,t_{1})\right>^2}, \label{mr2}
\eeq
is a solution of the Dym type equation (\ref{6.4.21k}).
Here $H^+$ and $F_{1}$ are given by
\beq
H^+&=& \frac{1}{2}\left<p,p\right>-\frac{1}{8}\left<q,q\right>
- \frac{1}{4\left<\La q,q\right>},\nonumber\\
F_{1}&=&-
\frac{\left<\La^2 q,q\right>}{4\left<\La q,q\right>}+
\frac{1}{2}\left(\left<\La q,p\right>^2-\left<\La q,q\right>\left<\La p,
p\right>\right).\nonumber
\eeq
\end{corollary}

By Theorem \ref{th4}, the Bargmann constraint given by Eq. (\ref{mch1})
is actually a solution of the positive order CH hierarchy (\ref{6.4.19}).
Thus, we also {\bf call the system $(H^+)$ (i.e. Eq. (\ref{H++}))
  a  positive order
constrained CH flow (i.e. Bargmann type) of the spectral problem (\ref{CHL1}).
  All Hamiltonian systems
$(F_k), \ k\ge 0,k\in\Z$ are therefore called the positive
order integrable Bargmann flows in the whole $\R^{2N}$}.

In a further procedure, we can get 
 the algebro-geometric solutions for the positive order
CH hierarchy.

\section*{Acknowledgments}

Z. Qiao would like to express his sincere thanks to
Dr. Camassa for his fruitful discussions and also to 
Dr. Gesztesy for his good suggestions.

This work was supported by the U.S. Department of Energy under
contracts W-7405-ENG-36 and the Applied Mathematical Sciences Program
KC-07-01-01;  the Special Grant of National
Excellent Doctorial Dissertation of China; and also the Doctoral
Programme Foundation of the Insitution of High Education of China.

\newpage
\section*{Appendix}
\begin{center} {\bf Abel mapping and the $\Theta $-function}
\end{center}
{\small
{\bf 1} \ \ If the genus of a Riemann surface is $g$, this surface is homomorphic
to a sphere with $g$ handles. Such a basic system of closed paths (or 
contours) $\alpha_1,...,\alpha_g,\beta_1,...,\beta_g$ can be chosen such
that the only intersections among them are those of $\alpha_i$ and 
$\beta_i$ with the same number $i$.

  Let the Riemann surface be covered with charts $\left(U_i,z_i\right)$, where 
$z_i$ are local parameters in open domains $U_i$, the transition from
 $z_i$ to $z_j$ in intersections $U_i \cap U_j$ being holomorphic. If 
in any $U_i$ a differential $\varphi _i \left(z_i\right)dz_i$ with meromorphic
 $\varphi _i \left(z_i\right)$ is given and in the common parts $U_i \cap U_j, \ 
 \varphi _i \left(z_i\right)dz_i = \varphi_j \left(z_j\right)dz_j$, then we say that there is an Abel differential
 $\Omega$ on the whole surface with restrictions $\Omega |_{v_i} =\varphi_i\left(z_i\right)dz_i$.
The Abel differential is of the first kind if all the $\varphi_i\left(z_i\right)$ are holomorphic.
There are exactly $g$ linearly independent differentials of the first kind $\omega_1,...,\omega_g$.
They are normed if $\int_{\alpha_i} \omega_j = \delta_{ij}$, which condition determines them 
uniquely. We shall always assume them normed. The numbers $\int_{\beta_i} \omega_j = B_{ij}$ are
 called $\beta$-periods. The matrix $B = \left(B_{ij}\right)$ has the following properties: 1) $B_{ij}=B_{ji}$,
 2) $\tau $= {\rm Im}  $B$ is a positive definite matrix.

  We consider a $g$-dimensional vector
  ${\cal A}\left(P\right) =\left\{\int_{P_0}^P \omega_j \right\}$, where
   $P_0$ is a fixed point of the Riemann surface and $P$ is an arbitrary point. This vector is
 not uniquely determined, but depends on the path of integration. If the latter is changed then
 a linear combination of $\alpha$ and $\beta$-periods with integer coefficients can be added: 
$\left({\cal A} \left(P\right)\right)_j \mapsto \left({\cal A}\left(P\right)\right)_j+\sum_{1}^{g} n_i \delta _{ij} +\sum_{1}^{g} m_i B_{ij}$,
i.e. ${\cal A}\left(P\right) \mapsto {\cal A}\left(P\right)+\sum n_i \delta_i+\sum m_i B_i$, where $\delta_i$ is the
 vector with coordinates $\delta_{ij},\ B_i$ is the vector with coordinates $B_{ij}$. Thus ${\cal A}
 \left(P\right)$ determines a mapping of the Riemann surface on the torus $\J=\C^g/\T$, where $\T$ is the lattice
 generated by $2g$ vectors $\{\delta_i, B_{i}\}$ (which are linearly independent over $\R$).
This mapping is called the Abel mapping, and the torus $\J$ is the Jacobi manifold or the Jacobian
 or the Riemann surface.

   The Abel mapping extends by linearity to the divisors:
   $${\cal A}\left(\sum n_k P_k\right)=\sum n_k {\cal A}\left(P_k\right).$$

$ $\\
{ \bf 2} \ \  $Abel\ \ Theorem$. Those and only those divisors go to zero of the Jacobian by the Abel
 mapping which are principal. The latter means that they are divisors of zeros and poles
of meromorphic functions on the surface. If $P_k$ is a zero of the function, then $n_k>0$ and
 $n_k$ is the degree of this zero. If $P_k$ is a pole, then $n_k<0$ and $|n_k|$ is the degree of
 this pole.

   Of special interest is the case of divisors of degree $g$ with all $n_k=1$,  
 i.e. of non-ordered
 sets of $g$ points $P_1,...,P_g$ of the Riemann surface. All the sets of such kind form the 
symmetrical $g$-th power of the Riemann surface.
The Abel mapping has the form
\beq
{\cal A}\left(P_1,...,P_g\right)
=\left\{ \sum_{j=1}^g\int_{P_0}^P \omega_j\right\},\  j=1,...,g. \nonumber
\eeq
The symmetrical $g$-th power is a complex manifold of the complex dimension $g$ and it is mapped
 on the Jacobian which is a manifold of the same dimension. A problem of conversion arises (the 
Jacobi inverse problem). It can be solved with the help of the $\Theta $-function.

$ $\\
{ \bf 3} \ \
    For arbitrary $P \in \C^g$, let
\beq
\Theta \left(P\right)&=&\sum_{Z\in \Z^g} {\rm exp} \left\{\pi i\left(BZ,Z\right)+2\pi i\left(P,Z\right)\right\}, \nonumber\\
         \left(BZ,Z\right)& =&\sum_{i,j=1}^g B_{ij} z_i z_j, \
         Z=(z_1,...,z_g)^T \nonumber \\
             \left(P,Z\right)&=&\sum_{i=1}^g p_i z_i,\
             P=(p_1,...,p_g)^T.
\nonumber 
\eeq
The series converges owing to the properties of the matrix $B$. The
$\Theta$-function has the
 properties
\beq
\Theta \left(-P\right)&=&\Theta \left(P\right); \nonumber \\ 
 \Theta \left(P+\delta_k\right)&=&\Theta \left(P\right); \nonumber \\
\Theta \left(P+B_k\right)&=&\Theta \left(P\right) {\rm exp} \left\{-\pi i\left(B_{kk}+2p_k\right)\right\}. \nonumber 
\eeq
Note that the $\Theta$-function is not defined on the Jacobian because of the latter property.

$ $\\
{\bf  4}\ \  $Riemann \ Theorem$. There are constants $\K=\{k_i\}, i=1,...,g$ (Riemann constants) determined
 by the Riemann surface such that the set of points $P_1,...,P_g$ is a
 solution of the system of
 equations
\beq
\sum_{i=1}^g \int_{P_0}^{P_i} \omega_j=l_j, \
\mathbb{L}=\{l_j\} \in \J, \ j=1,...,g, \nonumber 
\eeq
if and only if $P_1,...,P_g$ are the zeros of the function $\tilde{ \Theta }\left(P\right)= \Theta \left({\cal A}\left(P\right)
-\mathbb{L}-\K\right)$ (which has exactly $g$ zeros). Note that while the function $\tilde{ \Theta }\left(P\right)$ is not
 uniquely determined on the Riemann surface (it is multivalued) its zeros are
multivalued, since distinct 
branches of $\tilde{ \Theta }\left(P\right)$ differs by exponents.

$ $\\
{ \bf 5}\ \  We define now Abel differentials of the second and of the third kind. The Abel differential of the second kind, $\Omega_P^{\left(k\right)}, \ k=1,2,...,$ has the only singularity
 at the point $P$ which is a pole of the order $k+1$. The differential can be represented at this
 point as $dz_+^{-k}$ (holomorphic differential), $z$ being the local parameter at this point.
 Such a differential is uniquely determined if it normed: \
 $\int_{\alpha_i} \Omega_P^{\left(k\right)}=0, \ \forall i.$

   The Abel differential of the third kind $\Omega_{PQ}$ has only singularities which are simple
 poles at the points $P$ and $Q$ with the residues $+1$ and $-1$, respectively. It is uniquely 
determined by the same condition.

$ $\\
{ \bf 6} \ \ $Proposition$. If $z$ is a local parameter in a neighbourhood of the point $P$ and
$\omega_i=\varphi_i \left(z\right) dz$ is the Abel differential of the first kind, then
\beq
\frac{1}{2\pi i} \int_{\beta_i} \Omega_P^{\left(k\right)}
=\left.-\frac{1}{\left(k-1\right)!} \frac{d^{k-1}}{dz^{k-1}}
\varphi_i \left(z\right)\right|_{z=0}, \ i=1,...,g, \nonumber 
\eeq
and
\beq
\frac{1}{2\pi i} \int_{\beta_i}\Omega_{PQ}=\int_Q^P \omega_i, \ \ i=1,...g, \nonumber 
\eeq
which is also seen in Ref. \cite{DBA}.
}
                      
\end{document}